%% file: main.tex
\documentclass[tighten,twocolumn,trackchanges]{aastex7}
\usepackage{verbatim}
\usepackage{tabularx}
\usepackage{amsmath}
\usepackage{braket}
\usepackage{booktabs}
\usepackage{graphicx}
\usepackage{amssymb}
\usepackage{bm}	
\usepackage{threeparttable}
\usepackage{url}
\usepackage{newtxtext}
\usepackage{newtxmath}
\usepackage{xcolor}
\usepackage{hyperref}
\usepackage{appendix}

\newcommand{\kms}{\,km\,s$^{-1}$}
\newcommand{\mgii}{Mg\hspace{0.5mm}{\small II}}
\newcommand{\civ}{C\hspace{0.5mm}{\small IV}}
\newcommand{\hi}{H\hspace{0.5mm}{\small I}}
\newcommand{\ciii}{C\hspace{0.5mm}{\small III}}
\newcommand{\caii}{Ca\hspace{0.5mm}{\small II}}

\newcommand{\oi}{O\hspace{0.5mm}{\small I}}
\newcommand{\siiv}{Si\hspace{0.5mm}{\small IV}}

\newcommand{\nai}{Na\hspace{0.5mm}{\small I}}

\newcommand{\modify}[1]{\textcolor{black}{#1}}
\newcommand{\Ang}{\ifmmode\mathring{\mathrm{A}}\else\AA\fi}

\newcommand{\ionstage}[1]{\,\hspace{0.5mm}{\small\rmfamily #1}}
\shortauthors{Anand \& Muzahid}
\accepted{\today}

\begin{document}

\title{The First Blind Survey of Intervening \nai\ Absorbers: Incidence and Cosmic Evolution with DESI}

\shorttitle{Cosmic Evolution of Neutral Sodium Absorbers with DESI}
\include{authors}

\begin{abstract}
We present the first blind statistical survey of intervening \nai\ absorbers over the last $\simeq3$~Gyr using quasar spectra from the Dark Energy Spectroscopic Instrument Data Release~1. We identify 3636 \nai\ D doublets at $0.01\lesssim z_{\rm abs}\lesssim0.27$ along 214,035 quasar sightlines. We quantify the survey selection function through end-to-end injection--recovery tests and estimate the residual contamination from visual inspection. The completeness reaches 50\% at a rest-equivalent width of the \nai~$\lambda5891$ line, $W_{5891}\simeq0.95\,\Ang$, and the catalog-weighted purity is $\simeq77\%$. For the fiducial sample with $0.95<W_{5891}<3.5\,\Ang$, we measure the
equivalent-width distribution ($f(W_r)$), absorber incidence ($dN/dX$), and cosmic
\nai\ mass density ($\Omega_{\rm Na\,I}$).
From $z\simeq0.25$ to $z\simeq0.03$, $dN/dX$ rises by a factor of $\simeq3.4$ in both rest-equivalent-width subsamples, split at the median
$W_{5891}=1.41\,\Ang$, while $\Omega_{\rm Na\,I}$ increases by a similar
factor. This corresponds to a smooth increase in the product of the comoving number density of \nai-absorbing hosts ($n_{\rm com}$) and the cross-section of the \nai-bearing gas ($\sigma_{\rm eff}$) with cosmic time. The similar evolution of $\Omega_{\rm Na\,I}$ and $dN/dX$ indicates little change in the mean measured \nai\ column density per absorber. We also find preliminary evidence that \nai\ absorbers are associated with foreground galaxies at relatively small projected separations and that their quasar sightlines are redder than matched controls. These results support a picture in which \nai\ traces dense neutral gas in and around galaxies. Dedicated galaxy surveys and CGM/ISM simulations can test the inferred incidence and cross-section, while spectral stacking can probe the associated gas and dust.

\end{abstract}



\section{Introduction}\label{intro}

Absorption lines in quasar spectra provide a sensitive way to detect diffuse gas in and around galaxies and in the intergalactic medium (IGM), where the low gas density makes direct emission difficult to observe. Different ions trace different phases of this gas. In particular, low-ionization metal absorbers probe the cooler and more neutral phases of this gas. Neutral sodium (\nai) is especially sensitive to dense, well-shielded regions because its first ionization potential is
only 5.14~eV, well below the 13.6~eV threshold of neutral hydrogen. Sodium can therefore be ionized to \ion{Na}{2} even while \hi\ remains neutral, so \nai\ traces a distinct component of the \hi\ reservoir \citep{ferlet1985,benbekhti2008,churchill2025a}. The \nai\ $\lambda\lambda5891,5897$ doublet lies in the optical wavelength range at $z\lesssim0.7$, making it accessible to ground-based spectroscopic surveys.

For this reason, \nai\ absorption has been studied extensively in the Milky Way interstellar medium and halo, where it traces dense clouds and small-scale neutral structure \citep{ferlet1985,sembach1993,benbekhti2008}. Extragalactic \nai\ has also been detected in DLAs and sub-DLAs, in cool
outflows from star-forming and infrared-luminous galaxies
\citep{rupke2005a,rupke2005b,martin2006,richter2011}, and in low-luminosity active galactic nuclei \citep[AGNs;][]{moghni2026}. Studies of nearby galaxies show that the strength and detectability of \nai~absorption depend strongly on dust attenuation and star-formation surface density, highlighting the role of shielding in maintaining neutral sodium \citep{chen2010,avery2022}. Empirical correlations between \nai\ absorption, \hi\ column density, and dust reddening further support this connection, although these relations show substantial scatter due to ionization, depletion, saturation, and unresolved cloud structure \citep{sembach1993,poznanski2012}. This also makes \nai\ a potentially useful way to select candidate sub-DLA/DLA systems at low redshift from ground-based optical spectra, where direct \hi\ measurements from Ly$\alpha$ require ultraviolet space-based observations.

However, the cosmological population of \nai\ absorbers remains largely unexplored. Previous extragalactic studies have mainly searched for \nai~toward known galaxies, DLAs, \caii~absorbers, or galaxies with strong star formation, and \nai\ is detected only in a subset of neutral metal-bearing systems \citep{richter2011,rubin2022, churchill2025a,churchill2025b}. In contrast, blind surveys of species such as \mgii\ and \caii\ have established their incidence and equivalent-width distributions over large redshift ranges \citep{wild2005,nestor05,zhu13a,sardane2014}.
No comparable measurement exists for an absorption-selected \nai\
population, leaving the abundance and recent evolution of the cold ($T\sim 10^{2}-10^{3}\, \rm K$), well-shielded neutral phase traced by \nai\ not well established. To our knowledge, the large quasar sample and broad optical wavelength coverage of DESI \citep{DESI2022-Instrument,desidr1release2025} provide the first opportunity for a large blind census of intervening \nai\ absorbers at low redshift.

We use DESI DR1 quasar spectra to construct a uniformly selected \nai\ absorber catalog and quantify its completeness with injection--recovery simulations. Using the corrected sample, we measure the equivalent-width distribution, absorber incidence, and cosmic \nai\ mass density, allowing us to examine the recent evolution of the \nai-bearing neutral phase. We also present preliminary absorber--galaxy associations and reddening
measurements to explore the environments of these systems and provide initial constraints for comparison with larger observational samples and CGM/ISM simulations.

The paper is organized as follows. Section~\ref{desidata} describes the DESI DR1 quasar sample, and Section~\ref{methods} presents the analysis methods. Section~\ref{results} presents the absorber properties and statistical measurements. In Section~\ref{discussion}, we discuss their physical interpretation and preliminary host-galaxy results. We summarize the main results in Section~\ref{summary}. Throughout this work, we adopt a Planck cosmology \citep{planck20} with $\Omega_{\rm m}=0.3097$, $H_0=67.7$~\kms\,Mpc$^{-1}$, and $\Omega_{\Lambda}=1-\Omega_{\rm m}$.


\section{DESI Data}\label{desidata}

The Dark Energy Spectroscopic Instrument (DESI) is a fiber-based multiobject spectroscopic survey designed to map the three-dimensional distribution of galaxies and quasars over a large cosmological volume \citep{desidr1release2025}. The instrument is installed on the 4\,m Mayall Telescope at Kitt Peak National Observatory and uses 5000 robotically positioned fibers to obtain spectra simultaneously with 10 spectrographs \citep{DESI2016-Instrument}. The resolving power, $\mathcal{R}(\lambda)$, varies almost linearly from $2000$ at the blue end to $5500$ at the red end \citep{DESI2016-Instrument, DESI2022-Instrument}. The final flux-calibrated and sky-subtracted spectra cover $3600$--$9824\,\Ang$ on a uniform wavelength grid with a pixel size of $0.8\,\Ang$ \citep{guy2023}. This makes DESI spectra well suited for systematic searches for intervening absorption systems over a broad redshift range.

\begin{table*}
\centering
\begin{threeparttable}
\caption{Cumulative selection of the DESI DR1 quasar sample used
for the \nai\ absorber search.}
\label{tab:qso_selection}

\begin{tabular}{lr}
    \hline
    \hline
    Selection & Number remaining \\
    \hline
    
    \texttt{HEALPix}-based \texttt{agngal} catalog\tnote{a}, 
    \,\, $\texttt{AGN\_MASKBITS \& QN} \neq 0$\tnote{b}
    & 1,418,370 \\
    
    $\texttt{MAIN\_PRIMARY=True}$ and $\texttt{ZCAT\_PRIMARY=True}$\tnote{c}
    & 1,298,726 \\
    
    $\texttt{ZWARN}=0$, $\texttt{COADD\_FIBERSTATUS}=0$,
    $0.05 < z_{\rm qso} < 4.8$, and 
    $\texttt{FLUX\_{G,R,Z,W1,W2}} > 0$
    & 1,282,317 \\
    
    Non-BAL QSOs based on $\texttt{BI\_CIV}\leq0$ and $\texttt{AI\_CIV}\leq0$\tnote{d}
    & 1,153,296 \\
    
    Median per-pixel $\mathrm{S/N}_{\rm median,window} \geq 7$ in the \nai\ search window
    & 214,035 \\
    \hline
\end{tabular}

\begin{tablenotes}[flushleft]
    \footnotesize
    \item[a] DESI DR1 AGN/Galaxy VAC: \url{https://data.desi.lbl.gov/doc/releases/dr1/vac/agngal/}.
    \item[b] \texttt{AGN\_MASKBITS} is the AGN-selection bitmask in the VAC. We require the QuasarNet (\texttt{QN}) bit to be set, indicating that QuasarNet classifies the source as a QSO.
    \item[c] These flags denote main survey targets and entries from the primary redshift catalog.
    \item[d] BAL information is taken from the DESI DR1 Ly$\alpha$ Forest BAO catalog:
    \url{https://data.desi.lbl.gov/doc/releases/dr1/vac/zlya/}. QSOs matched to this catalog and having \texttt{BI\_CIV > 0} or \texttt{AI\_CIV > 0} are removed. Unmatched lower-redshift QSOs
    ($z_{\rm QSO}<1.6$), for which CIV~BAL measurements are not available,
    are retained.
\end{tablenotes}

\end{threeparttable}
\end{table*}

\subsection{DR1 Quasar Sample}
\label{sec:qso_sample}

We use quasar spectra from DESI Data Release~1
\citep{desidr1release2025}. The construction of the parent quasar sample is summarized in Table~\ref{tab:qso_selection}. Briefly, we start from the HEALPix-based DR1 \texttt{agngal} value-added catalog and apply a series of classification and quality cuts, supplemented with information from the DESI DR1 Broad Absorption Line (BAL) catalog. The quasar redshifts are based on a combination of the PCA-based \texttt{redrock} pipeline \citep{anand24,bailey2026} and quasar afterburners using emission-line and CNN-based classification methods \citep{busca18,chaussidon2023}.

For the final \nai\ parent sample, we further require the median per-pixel signal-to-noise ratio within the \nai\ search window to satisfy ${\rm S/N}_{\rm median,window}\geq7$. Injection-recovery tests show that the completeness rises rapidly with this quantity and reaches $\simeq50\%$ at ${\rm S/N}_{\rm median,window}\simeq7$ (Appendix~\ref{appendix:selection_function}). This threshold removes low-quality sightlines that would require large completeness corrections, and defines a high-quality search path
for the statistical analysis. After applying all selection criteria, the final parent sample contains $214,035$ unique quasars to be searched for \nai~absorbers.

\begin{figure}
    \centering
    \includegraphics[width=0.975\linewidth]{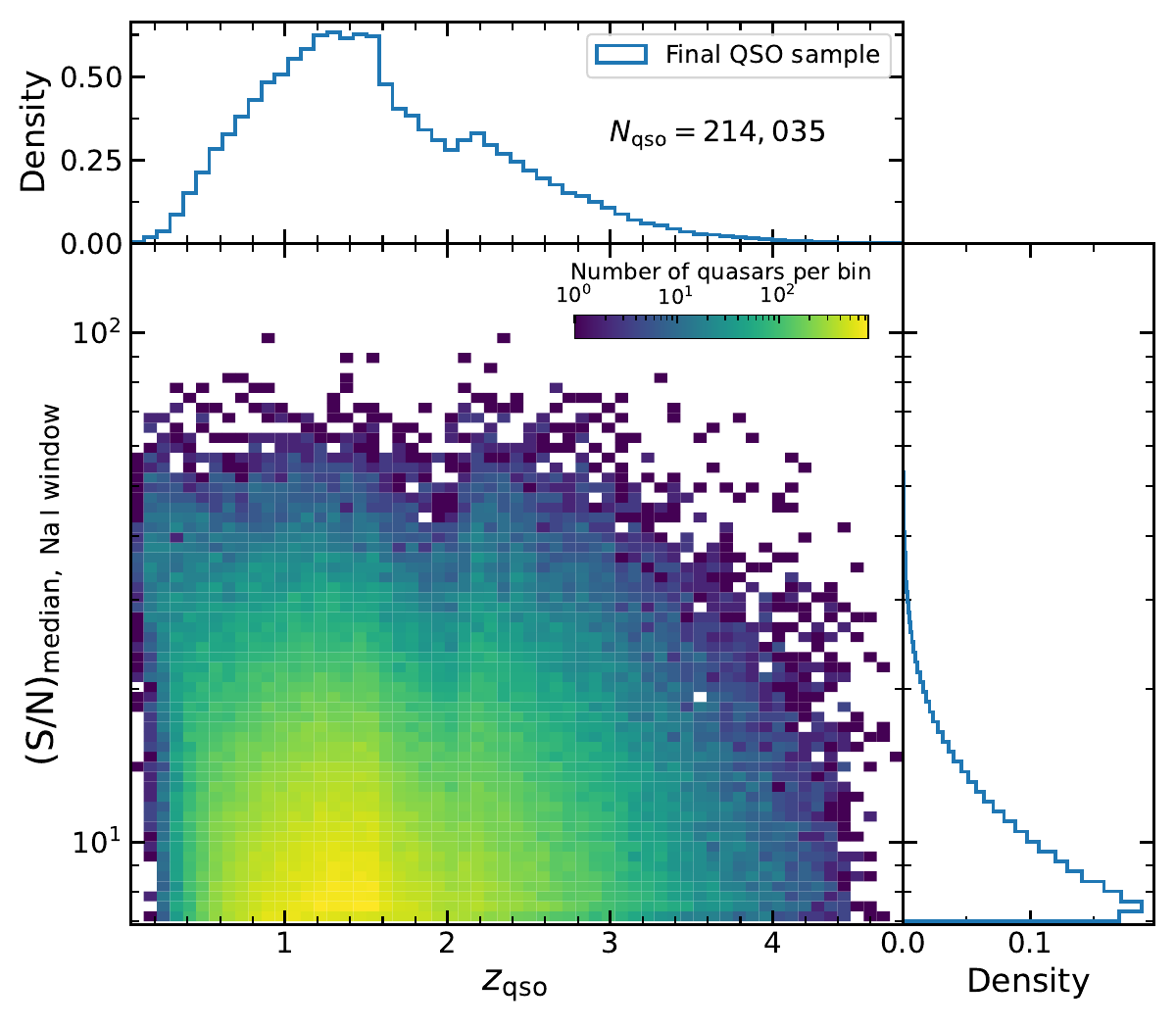}
    \caption{Redshift and signal-to-noise distributions of the final quasar sample used for the \nai\ absorber search. The main panel shows the two-dimensional distribution of $z_{\rm qso}$ and the median S/N within the \nai\ search window, with the color scale indicating the number of quasars per bin. The upper and right panels show the corresponding normalized marginal distributions.}
    \label{fig:qso_sample}
\end{figure}

Figure~\ref{fig:qso_sample} shows the redshift distribution and median signal-to-noise properties of the final QSO sample used for the \nai\ absorber search. The redshift distribution shows features near $z_{\rm QSO}\simeq1.6$ and $z_{\rm QSO}\simeq2.1$. The feature near $z_{\rm QSO}\simeq1.6$ is introduced by the \civ\ BAL-removal step, since BAL classifications are available only where \civ\ can be measured. The feature near $z_{\rm QSO}\simeq2.1$ is associated with the prioritization of Ly$\alpha$ quasars in DESI for BAO measurements.

\section{Methods}\label{methods}

\subsection{Quasar Continuum Fitting}
\label{sec:continuum}

We estimate the quasar continua with \texttt{nmfqsofit}\footnote{\url{https://github.com/abhi0395/nmfqsofit/}}, a code originally developed for continuum fitting of SDSS DR16 quasars  \citep{anand2021}. The method is based on a non-negative matrix factorization \citep[NMF;][]{lee1999} representation of quasar spectra, following the approach applied to SDSS DR7 quasars by \citet{zhu13a}. For this work, we adapted the framework to the DESI wavelength sampling, spectral uncertainties, masks, and data format. We briefly summarize the key mathematical steps of the NMF approach below. 

The continuum, $C_{\rm NMF} (\lambda)$ of each quasar is represented as a non-negative linear combination of the NMF eigenspectra,

\begin{equation}
    C_{\mathrm{NMF}}(\lambda)
    =
    \sum_{k=1}^{N_{\mathrm{comp}}}
    a_k E_k(\lambda),
    \qquad a_k\geq0
    \label{eq:nmf_continuum}
\end{equation}
where $E_k(\lambda)$ is the $k$th eigenspectrum, $a_k$ is its coefficient, and $N_{\rm comp}$ is the total number of NMF eigenspectra. The coefficients are determined using a non-negative least-squares solver by minimizing the inverse-variance-weighted statistic
\begin{equation}
    \chi^2
    =
    \sum_{i\in\mathcal{V}}
    \frac{
    \left[F_i-C_{\mathrm{NMF}}(\lambda_i)\right]^2
    }{
    \sigma_i^2
    }
    \label{eq:nmf_chi2}
\end{equation}
where $F_i$ and $\sigma_i$ are the observed flux and its
uncertainty, and $\mathcal{V}$ denotes the valid pixels included in the fit. Pixels masked by the DESI spectral masks are excluded, while absorption-like pixels ($3\sigma$ negative deviations only) are iteratively masked to prevent strong absorption features from biasing the continuum low.

\subsubsection{Construction of the DESI eigenspectra}
\label{sec:eigenspectra}

The NMF eigenspectra used in this work were constructed from DESI DR1 quasars following the procedure described in \citet{anand2021}, which we briefly summarize here. We construct the eigenspectra independently in four overlapping quasar-redshift bins: $0<z_{\rm qso}<1.2$, $0.6<z_{\rm qso}<2.1$,
$1.6<z_{\rm qso}<3.3$, and $2.6<z_{\rm qso}<5.0$. In each interval, we use $100,000$ randomly selected non-BAL DESI DR1 quasars while preserving the parent redshift and magnitude distributions. Their spectra are normalized by the mean flux within the rest-frame windows 4150--4250, 3018--3100, 2130--2240, and 1420--1500~$\Ang$, respectively, for the corresponding redshift ranges. These windows provide adequate wavelength coverage while avoiding prominent
quasar emission features. We construct $N_{\rm comp}=12$ eigenspectra in each redshift interval, as in \citet{anand2021}, which we find sufficient to model the SDSS or DESI-like quasar spectra. Because the eigenspectra are constructed from normalized spectra, each quasar spectrum is normalized before fitting. The reconstructed continuum is then rescaled to the original flux level using the same normalization factor.

For quasars covered by more than one set of eigenspectra, \texttt{nmfqsofit} fits each valid set and selects the solution with the largest number of usable pixels within its rest-frame wavelength coverage. We use this approach over minimum $\chi^2$, because a fit that partly follows absorption features can artificially reduce the $\chi^2$.

\subsubsection{Continuum correction and residual spectrum}
\label{sec:continuum_correction}

Following the initial NMF reconstruction, we correct residual intermediate- and small-scale continuum variations using the iterative median-filtering procedure described in \citet{anand2021}. Briefly, absorption-like pixels are masked using $1.5\sigma$ clipping (negative deviations only), and the ratio of the observed flux to the NMF continuum is smoothed sequentially using 141- and 71-pixel median-filter kernels. These kernel sizes are chosen to be much broader than typical absorption features, minimizing the risk of fitting out genuine absorption. The final continuum, $C_{\rm final}(\lambda)$, is the product of the initial NMF continuum and the smooth median-filter correction. The smooth corrected continuum is retained only when it decreases the $\chi^2$ statistic in  Eqn.~\ref{eq:nmf_chi2}, for the model and observed spectrum. The final normalized spectrum used for the absorber search and its pixel-wise uncertainty are
\begin{equation}
    R(\lambda)
    =
    \frac{F(\lambda)}{C_{\mathrm{final}}(\lambda)},
    \qquad
    \sigma_R(\lambda)
    =
    \frac{\sigma_F(\lambda)}
         {C_{\mathrm{final}}(\lambda)}
    \label{eq:final_residual}
\end{equation}
The quoted uncertainty propagates only the spectral flux error and does not include uncertainty in the fitted continuum.

We use \texttt{nmfqsofit} version
\texttt{v1.1.1}\footnote{\url{https://github.com/abhi0395/nmfqsofit/releases/tag/v1.1.1}} with the DESI DR1 eigenspectra released as \texttt{nmfeigenspectra} version
\texttt{v1.0.0}\footnote{\url{https://github.com/abhi0395/nmfeigenspectra/releases/tag/v1.0.0}}. The implementation is parallelized for efficient application to the full DESI quasar sample. We show an example spectrum with the resulting continuum fit and normalized
residual in the top panel of Figure~\ref{fig:spectra}.

\subsection{Automated Detection of \nai~Absorbers}
\label{sec:qsoabsfind}

We search the continuum-normalized quasar spectra for intervening \nai~absorption systems using \texttt{qsoabsfind}, an automated Python package for identifying resonance doublets in low-resolution spectra. The method was originally developed for \mgii~searches in SDSS quasars \citep{anand2021,anand2022} and was subsequently extended to \civ~absorbers in DESI spectra \citep{anand2025}. We use version \texttt{2.1.0} of \texttt{qsoabsfind}\footnote{\url{https://github.com/abhi0395/qsoabsfind/releases/tag/v2.1.0}}, which provides a generic framework for identifying absorption-line doublets in low-resolution SDSS- and DESI-like spectra.

The pipeline combines matched-kernel convolution, line-significance thresholds, doublet-profile fitting, and selection criteria based on the expected wavelength separation and relative physical properties of the two transitions. The implementation uses multiprocessing and is designed for survey-scale searches on high performance computing clusters. 

\subsubsection{Absorber search window}
\label{sec:nai_search_window}

For each quasar, we define the observed wavelength interval over which intervening \nai\ absorbers can be detected. The blue boundary ($\lambda_{\rm blue, \, obs}$) is set by the quasar Ly$\alpha$ emission wavelength to avoid contamination from the Ly$\alpha$ forest, while the red boundary ($\lambda_{\rm red, \, obs}$) is set by the redder component of the \nai\ doublet ($\lambda5897$) at the quasar redshift. We additionally exclude absorbers within $\Delta v=4000~\mathrm{km\,s^{-1}}$ of the quasar to reduce contamination from associated systems and quasar outflows. Therefore,

\begin{equation}
    \begin{split}
    \lambda_{\rm blue,obs}
    &=
    \lambda_{\rm Ly\alpha}
    \left(1+z_{\rm qso}+\Delta z_v\right)\\
    \lambda_{\rm red,obs}
    &=
    \lambda_{\rm NaI,red}
    \left(1+z_{\rm qso}-\Delta z_v\right)
    \end{split}
    \label{eq:nai_qso_window}
\end{equation}
where
$\Delta z_v=(|\Delta v|/c)(1+z_{\rm qso})$. Using these our final search window is
\begin{equation}
    \begin{split}
    \lambda_{\rm start}
    &=
    \max\left(
    \lambda_{\rm obs,min},
    \lambda_{\rm blue,obs},
    5955~\Ang
    \right)
    +8~\Ang\\
    \lambda_{\rm end}
    &=
    \min\left(
    \lambda_{\rm obs,max},
    \lambda_{\rm red,obs},
    7500~\Ang
    \right)
    -8~\Ang
    \end{split}
    \label{eq:nai_search_window}
\end{equation}
where $\lambda_{\rm obs,min}=3600\,\rm \Ang$ and $\lambda_{\rm obs,max}=9824\,\rm \Ang$ are the wavelength limits of the DESI spectra. The $8~\Ang$ offset was applied so that the doublet components are slightly away from the spectral edges. The lower limit ($5955\,\rm \Ang$) corresponds to $z_{\rm abs}\gtrsim0.01$ and excludes Galactic and atmospheric \nai, while the upper limit ($7500\,\rm \Ang$) exlcudes the lower-efficiency DESI $z$-camera region and its stronger sky-subtraction residuals.

We exclude candidate systems for which either member of the \nai\ doublet falls within $\pm2000~\mathrm{km\,s^{-1}}$ of prominent QSO-frame emission lines, including \siiv, \civ, \ciii, and \mgii. These regions are removed because associated absorption or line-confusion near the quasar emission redshift can mimic intervening \nai\ doublets.

We also mask $\pm 2000$ \kms~regions around strong narrow emission lines, including [Ne\ionstage{III}], [Ne\ionstage{V}], [N\ionstage{II}], [O\ionstage{I}], [O\ionstage{II}], H$\alpha$, H$\beta$, H$\gamma$, and [O\ionstage{III}], where continuum mismatches can produce absorption-like residuals. Finally,
we mask the observed-frame interval $6295$--$6305~\Ang$ to avoid contamination from [\oi] $\lambda6300$ skyline.

\subsubsection{Candidate identification}
\label{sec:nai_candidates}

The continnum normalized  spectrum is convolved with a double-Gaussian kernel matching the separation of the \nai\ $\lambda\lambda5891,5897$ doublet. We use Gaussian widths of 2--6 DESI pixels to retain sensitivity to systems with different observed line widths.

At each pixel, the local noise of the convolved spectrum is estimated within a 50-pixel window. Pixels exceeding a local $2\sigma$ significance threshold are identified as the first set of candidates, allowing the threshold to adapt to variations in spectral noise and residual quality. Contiguous candidate pixels are grouped into individual absorption systems following \citet{anand2021}. The absorber redshift of each group is taken as the weighted mean of the candidate-pixel redshifts, with weights proportional to the cube of the absorption depth ($1-R_i$), so that pixels closest to the line center receive the largest weight. Candidates recovered with different kernel widths are then merged, and duplicate detections are removed before profile
fitting.

\subsubsection{Doublet fitting and absorber redshifts}
\label{sec:nai_fitting}

Each candidate is then fitted with a double-Gaussian absorption model, with the amplitude, centroid, and observed Gaussian width of each component allowed to vary independently. The fit is weighted by the uncertainties in the normalized spectrum spectrum. The redshift inferred from each component is
\begin{equation}
    z_j
    =
    \frac{\lambda_{c,j}}{\lambda_{0,j}}-1
    \label{eq:nai_line_redshift}
\end{equation}
where $\lambda_{c,j}$ and $\lambda_{0,j}$ are its observed-frame fitted centroid and vacuum rest wavelength, respectively. The final absorber redshift is taken as their inverse-variance-weighted mean.

We retain only candidates whose fitted rest-frame doublet
separation ($\Delta \lambda_{\rm fit}$)\footnote{$\Delta \lambda_{\rm fit} = \frac{\lambda_{c, 5897} - \lambda_{c, 5891}}{1+z_{\rm abs}}$, where $\lambda_{c, 5897}$ and $\lambda_{c, 5891}$ are fitted line centroid of the doublet in observed frame.} satisfies
\begin{equation}
    \left|
    \Delta\lambda_{\rm fit}
    -
    \Delta\lambda_{\rm NaI}
    \right|
    \leq 0.6~\Ang
    \label{eq:nai_line_separation}
\end{equation}
This criterion rejects unrelated features that do not reproduce the expected \nai\ doublet separation. The true rest-frame doublet separation for \nai~is $\Delta \lambda_{\rm NaI} = 5.98\,\rm \Ang$.

\subsubsection{Equivalent widths and line widths}
\label{sec:nai_measurements}

The rest-frame equivalent width of each component is measured directly from the normalized spectrum by trapezoidal integration over $\pm3\sigma_{\lambda,j}^{\rm obs}$ around its fitted observed-frame centroid,
\begin{equation}
    W_{r,j}
    =
    \frac{1}{1+z_{\rm abs}}
    \int_{\lambda_{c,j}^{\rm obs}-3\sigma_{\lambda,j}^{\rm obs}}
         ^{\lambda_{c,j}^{\rm obs}+3\sigma_{\lambda,j}^{\rm obs}}
    \left[1-R(\lambda_{\rm obs})\right]
    \,{\rm d}\lambda_{\rm obs}
    \label{eq:nai_ew}
\end{equation}
where $\lambda_{c,j}^{\rm obs}$ and
$\sigma_{\lambda,j}^{\rm obs}$ are the fitted observed-frame centroid and Gaussian width of the $j$th doublet component, respectively. The factor $(1+z_{\rm abs})^{-1}$ converts the observed-frame equivalent width to the absorber rest frame. If the two integration intervals overlap, they are truncated at the midpoint between the fitted observed-frame centroids to avoid assigning the same absorbed flux to both components.

The equivalent-width uncertainty is propagated over the same integration window using the normalized error spectrum and trapezoidal quadrature weights. This uncertainty is statistical; the additional continuum-placement term is expected to be subdominant for the strong systems used in the statistical sample.

We also report velocity dispersions after subtracting the DESI instrumental broadening in quadrature from the fitted Gaussian width,
\begin{equation}
    \sigma_{v,\mathrm{int}}
    =
    \left(
    \sigma_{v,\mathrm{fit}}^2
    -
    \sigma_{v,\mathrm{inst}}^2
    \right)^{1/2}, \,\,
    \sigma_{v,\mathrm{inst}}
    =
    \frac{c}{2.355\,\mathcal{R}(\lambda_{\rm obs})}.
    \label{eq:nai_velocity_width}
\end{equation}
Here \(\mathcal{R}(\lambda_{\rm obs})\) is the DESI resolving power at the observed wavelength of the transition \footnote{We approximate $\mathcal{R}(\lambda_{\rm obs})$ by linearly interpolating between the adopted
blue- and red-wavelength resolving powers, i.e., $\mathcal{R}=2000$ at
$\lambda_{\rm obs}=3600\,\Ang$ and $\mathcal{R}=5500$ at
$\lambda_{\rm obs}=9824\,\Ang$.}. We require \(\sigma_{v,\mathrm{fit}}^2 > \sigma_{v,\mathrm{inst}}^2\) for both doublet members. This is a quality-control cut on the fitted profile, not a statement that the intrinsic \nai\ components are resolved by DESI. It removes sub-instrumental fitted features, which are more likely to arise from noise spikes, bad pixels, or narrow continuum residuals. In the fiducial sample, the accepted systems have a median \(\sigma_{v,\mathrm{fit}}/\sigma_{v,\mathrm{inst}}\simeq1.8\), with 16th--84th percentiles of \(1.3\)--\(2.8\), and only \(\simeq2\%\) lie within 5\% of the rejection threshold. The selected sample therefore does not show a strong accumulation near this boundary. We do not impose any additional hard cut on the difference between the intrinsic velocity widths of the two doublet members.

\begin{figure*}
    \centering
    \includegraphics[width=\textwidth]{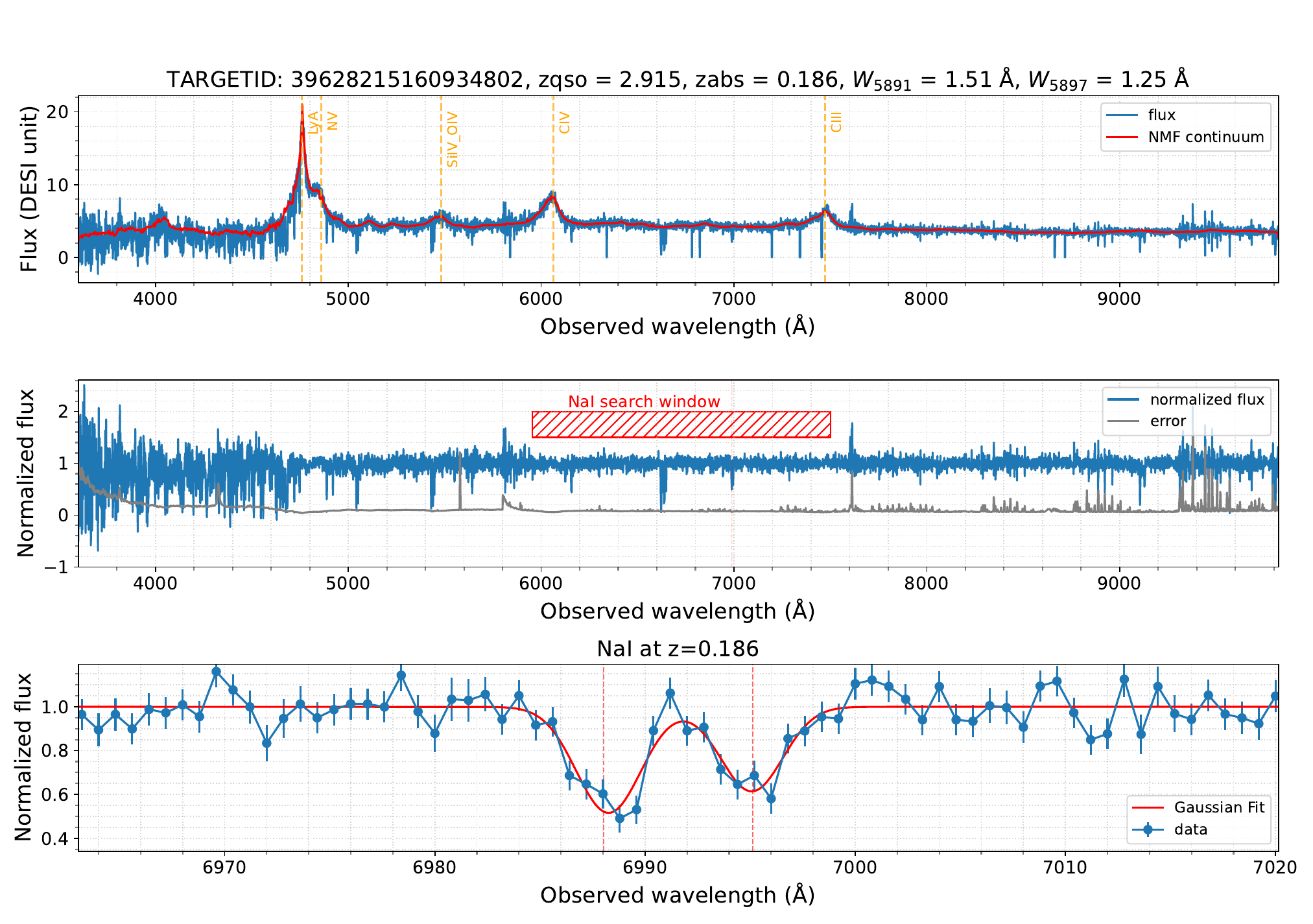}
    \caption{
    Example DESI quasar spectrum illustrating the continuum fitting and \nai~absorber detection. \textbf{Top}: observed-frame quasar flux (black) and the best-fitting \texttt{nmfqsofit} continuum (red), where flux and continuum are in units of $\rm 10^{-17}\, erg\, cm^{-2}\, s^{-1}\, \Ang^{-1} $. The blue region ($\lambda_{\rm obs}<5000\,\rm \Ang$) lies in the $\rm Ly\alpha$ forest and is excluded from the search. \textbf{Middle}: continuum-normalized spectrum spectrum and its corresponding uncertainty in the observed frame, showing a detected \nai~at $z_{\rm abs}= 0.186$. The hatched rectangle shows the \nai~wavelength search window. \textbf{Bottom}: zoom-in around a detected \nai~$\lambda\lambda5891,5897$ system with the red curve showing the best-fitting double-Gaussian model. Vertical dashed lines mark the expected positions of the two doublet components at the fitted absorber redshift. The measured rest-frame equivalent widths are $\rm W_{\rm 5891}=1.51\,\Ang$ and $\rm W_{\rm 5897}=1.25\,\Ang$. }
    \label{fig:spectra}
\end{figure*}

\subsubsection{Final candidate selection}
\label{sec:nai_selection}

Next, we define the doublet ratio (DR) as the ratio of the rest-equivalent
widths of the $\lambda5891$ and $\lambda5897$ lines, denoted by
$W_{5891}$ and $W_{5897}$, respectively:
\begin{equation}
    {\rm DR}
    =
    \frac{W_{5891}}{W_{5897}} .
    \label{eq:nai_dr}
\end{equation}
For a physical resonance doublet, \({\rm DR}\) is expected to lie between approximately one for saturated absorption and two in the optically thin limit. Allowing for measurement uncertainty, we require \(1-\sigma_{\rm DR} \leq {\rm DR} \leq 2+\sigma_{\rm DR}\), where \(\sigma_{\rm DR}\) is computed by standard error propagation from the equivalent-width uncertainties of the two doublet members\footnote{We compute \(\sigma_{\rm DR}={\rm DR} [(\sigma_{W_{5891}}/W_{5891})^2+ (\sigma_{W_{5897}}/W_{5897})^2]^{1/2}\),
assuming independent equivalent-width errors.}.

Finally, we also measure the local absorption signal-to-noise ratio of each doublet component over the fitted absorption window, using the summed absorption depth relative to the propagated normalized-flux uncertainty. The final selection requires that the both members satisfy the \textit{\nai~S/N criteria}
\begin{equation}
    {\rm S/N}_{5891} \geq 3,
    \qquad
    {\rm S/N}_{5897} \geq 2 
    \label{eq:nai_final_selection}
\end{equation}

Together with the doublet-separation and line-width criteria defined above, these cuts reduce contamination significantly which may arise from noise, continuum residuals, unrelated absorption features, and sky-subtraction residuals. Their effect on the detection completeness is quantified using the injection-and-recovery simulations described in Section~\ref{completeness}.

For each accepted system, \texttt{qsoabsfind} also measures the fitted line properties, equivalent widths, signal-to-noise ratios, doublet ratios, and their corresponding uncertainties. The bottom two panels of Figure~\ref{fig:spectra} show an example of \modify{a} detected \nai\ system and the corresponding double-Gaussian fit.

\subsection{\nai~Column Densities}
\label{column_density}

Because the intrinsic velocity components of the \nai\ absorbers are generally unresolved at DESI resolution, we cannot measure their true column densities directly. We therefore estimate \nai\ column densities using the apparent optical depth method (AODM; \citealt{savage1991, jenkins1996}), following the implementation
described in \citet{anand2025}. The method provides lower limits for strongly saturated systems and allows a saturation correction in less severe cases. For each detected system in our catalog, the AOD profile of each doublet component is integrated over $\pm150~\mathrm{km\,s^{-1}}$ around each line center. This window is approximately half the \nai\ doublet separation and is chosen to include most of the absorption while minimizing contamination from the companion transition.

The two transitions are combined using an AODM-based classification scheme. When both transitions provide valid measurements and show no significant AOD evidence for unresolved saturation, we adopt the inverse-variance weighted mean of the two integrated column densities. Unresolved saturation is diagnosed from the difference between the velocity-integrated AOD columns of the weak and strong transitions. When the weak transition gives a significantly larger AOD column than the strong transition, we apply the saturation correction of \citet{savage1991} and \citet{jenkins1996} when the system lies within their calibrated regime. More severe cases are retained as lower limits. If only one transition provides a usable measurement, the corresponding single-line AODM column is retained, with the saturation state flagged as indeterminate. Cases where the weak transition gives a significantly smaller column than the strong transition are flagged as inconsistent doublets.

Pixels at or below the adopted AODM flux floor ($\rm AODM_{floor}= 0.005 $) are retained at the floor and flagged as saturated or lower-limit pixels. Flux values above unity are not clipped, so that noise is not treated asymmetrically. Each system is assigned a column-density flag, $f_N$, which records how
the reported AODM column density was obtained: an inverse-variance weighted doublet measurement ($f_N=1$), a single-line measurement ($f_N=2,3$), a saturation-corrected weak-line estimate ($f_N=4$), a lower limit ($f_N=5,6$), an inconsistent doublet ($f_N=7$), or a failed measurement ($f_N=-1$).

The measured column densities span $12.05\lesssim\log[\mathcal{N}(\mathrm{Na\,I})/\mathrm{cm}^{-2}]\lesssim13.8$, with the median value of $\log[\mathcal{N}(\mathrm{Na\,I})/\mathrm{cm}^{-2}] = 12.80$, and the median column-density uncertainty is $\sigma_{\log\mathcal{N}}\simeq0.23$.

\subsection{Catalog Completeness and Purity}
\label{completeness}

\begin{figure*}
    \centering
    \begin{minipage}{0.55\textwidth}
        \centering
        \includegraphics[width=0.99\linewidth]
        {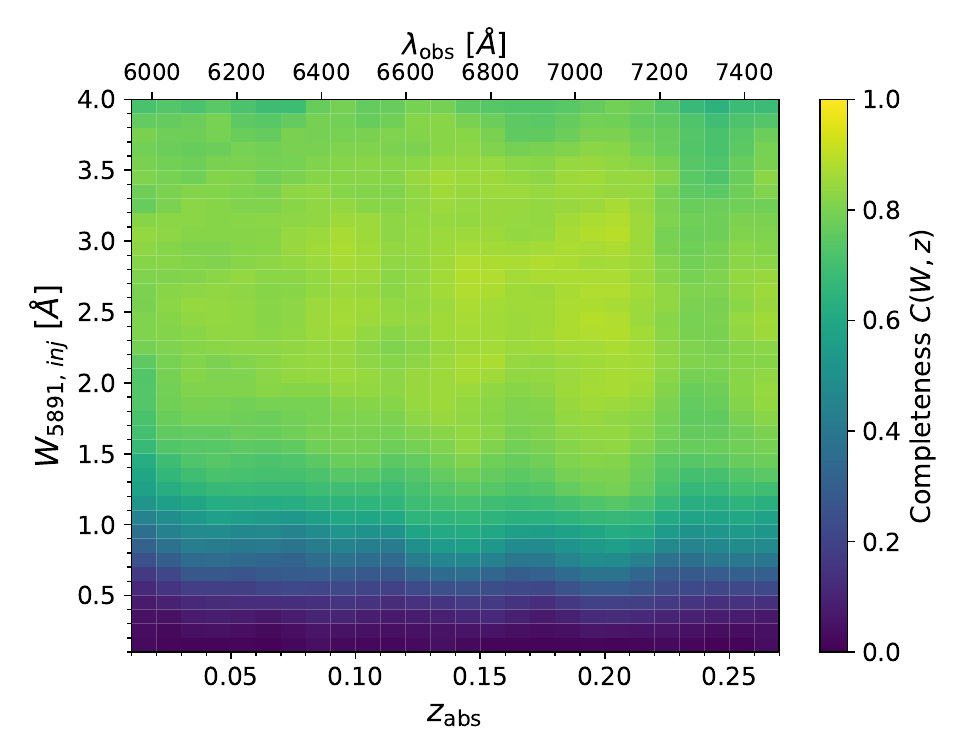}
    \end{minipage}
    \begin{minipage}{0.4\textwidth}
        \centering
        \includegraphics[width=0.9\linewidth]
        {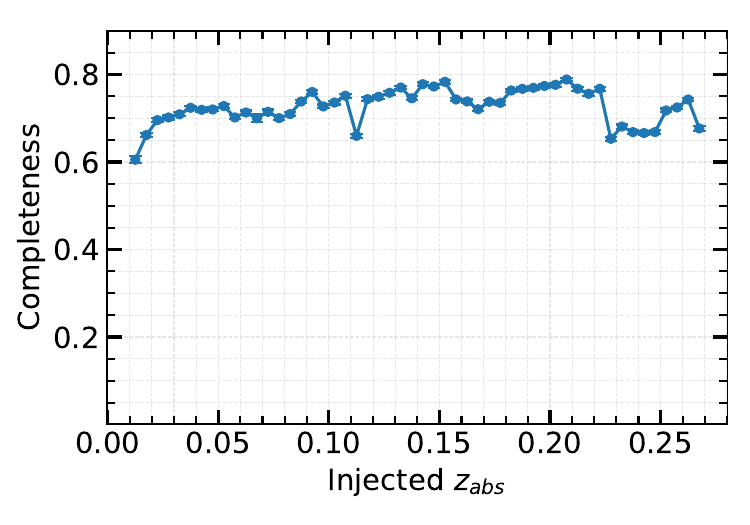}
        \includegraphics[width=0.9\linewidth]
        {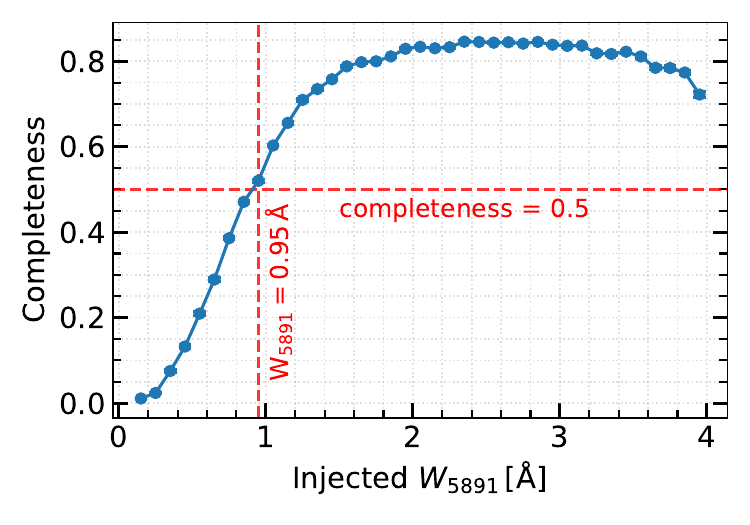}
    \end{minipage}
    \caption{
    Completeness of the \nai\ absorber search estimated from injection-and-recovery simulations.
    \textbf{Left:} Two-dimensional completeness as a function of absorber redshift and $W_{5891}$.
    \textbf{Top right:} Completeness marginalized over \modify{redshift}.
    \textbf{Bottom right:} Completeness marginalized over \modify{equivalent width}. In the right panels, the one-dimensional measurements are also marginalized over the injected doublet ratios, amplitudes, and line widths.
    The completeness is quantified using the complete continuum-fitting and absorber-detection pipeline.
}
    \label{fig:completeness}
\end{figure*}

We quantify the detection completeness by injecting synthetic \nai\ doublets into real DESI quasar spectra and repeating the complete catalog-construction procedure. The same median-S/N requirement is applied to the injected absorbers when constructing the completeness map, so that the measured completeness corresponds directly to the final statistical sample. The stronger-line equivalent width and Gaussian absorption amplitude are drawn uniformly over $0.1 \leq W_{5891} \leq 4.0~\Ang$ and $0.05<A_{5891}<0.95$, respectively. The doublet ratio is sampled uniformly over $1\leq{\rm DR}\leq2$, and the Gaussian width is determined from the sampled equivalent width and amplitude,
with the resulting injected profiles spanning approximately $\sigma_v\simeq30$--$130~{\rm km\,s^{-1}}$, covering the bulk of the observed width distribution. We additionally require the injected width to be at least as large as the wavelength-dependent instrumental width, $\sigma_v\simeq30$--$38~{\rm km\,s^{-1}}$, defined in Section~\ref{sec:nai_measurements}.

For each trial, the absorber redshift is selected randomly from the valid search path of a real quasar, including the same wavelength boundaries and masks used in the blind search. Synthetic \nai\ doublets are inserted multiplicatively into the observed DESI flux spectra using Gaussian optical-depth profiles for the two doublet
members. After injection, we refit the continuum with \texttt{nmfqsofit} and rerun the full \texttt{qsoabsfind} algorithm on the continuum-normalized spectrums, using the same search parameters and final selection criteria as for the real catalog. An injected absorber is considered recovered if the detected candidate is matched to the input absorber within $|\Delta v|\leq100~\mathrm{km\,s^{-1}}$ and the fractional equivalent-width error for each doublet member, $|W_{r,j}^{\rm inj}-W_{r,j}^{\rm rec}|/W_{r,j}^{\rm inj}$, is less than or equal to 0.5 for $j=5891,5897$. The adopted matching criteria are conservative: 99\% of detected injections are recovered within $|\Delta v|\leq50~\mathrm{km\,s^{-1}}$, and 94\%
have recovered equivalent widths within 50\% of the injected values for both doublet members. We perform approximately $6.5\times10^5$ trials, injecting at least three absorbers into each parent quasar spectrum.

The two-dimensional completeness in bins of absorber redshift ($z_{k}$) and equivalent width, $W_{k, \rm 5891}$ is then defined as,
\begin{equation}
    C(W_{k},z_{k})
    =
    \frac{N_{\rm rec}(W_{k},z_{k})}
         {N_{\rm inj}(W_{k},z_{k})}
    \label{eq:nai_completeness}
\end{equation}
where $N_{\rm inj}$ and $N_{\rm rec}$ are the numbers of injected and recovered systems, respectively. The completeness map is evaluated on a discrete grid in $W_{5891}$ ($\Delta W=0.1\,\rm \Ang$) and $z_{\rm abs}$ ($\Delta z = 0.01$). For each $i^{\rm th}$ absorber, the completeness correction $C_i=C(W_k,z_k)$ is assigned from the two-dimensional $k^{\rm th}$ bin whose centre is closest to the measured equivalent width and redshift of that absorber. No interpolation between adjacent bins is applied. Owing to the large simulation sample, each completeness bin contains sufficiently many trials that the corresponding binomial uncertainties are small. They are smaller than the plotting symbols in the one-dimensional marginalized distributions and are therefore not shown.

As shown in Figure~\ref{fig:completeness}, the completeness depends primarily on absorber strength. It rises from approximately 50\% at $W_{5891}\simeq 0.95~\Ang$ to about 85\% near $W_{5891}\simeq 1.9$--$2.0~\Ang$, and remains roughly constant at this level up to $W_{5891}\simeq 3.5~\Ang$. At the highest equivalent widths, the completeness decreases mildly. This is likely because these systems are often broader and more complex, making the doublet centroid and component structure harder to recover accurately with the automated search pipeline. The two-dimensional $(W_{5891}, z_{\rm abs})$ completeness map shows the same overall behavior: the recovery rate is mainly driven by equivalent width, with only a weaker dependence on absorber redshift across the surveyed redshift range. The redshift dependence at fixed equivalent width is comparatively modest, although variations remain because of wavelength-dependent noise, resolution, and masking.

We also estimate catalog contamination empirically through visual inspection of a stratified subset of candidates selected in bins of absorber redshift and equivalent width. In total, we inspected $\sim 600$ candidates, corresponding to $\simeq17\%$ of the final catalog. The sampling was designed to cover the full catalog while providing sufficient visual statistics in the equivalent-width range used for the statistical measurements. Each candidate was classified as GOOD, MAYBE, or BAD, corresponding to convincing \nai\ systems, plausible but uncertain systems, and likely spurious or misidentified systems, respectively. We assign visual scores of 1.0, 0.75, and 0.0 to these classes. The intermediate score for MAYBE reflects that these candidates remain plausible absorbers, although their classification is less certain.

For the statistical measurements, each $i^{\rm th}$ absorber is assigned a weight
\begin{equation}
    w_i =
    \frac{P_{{\rm visual},j}}{C(W_k,z_k)},
\label{eq:nai_weight}
\end{equation}
where $(W_k,z_k)$ denotes the centre of the completeness-map bin assigned to the $i^{\rm th}$ absorber, and $j$ denotes the corresponding stratified bin used for the visual-purity estimate. The completeness factor $C(W_k,z_k)$ is taken from the injection-recovery completeness map. The quantity $P_{{\rm visual},j}$ is the visual-purity factor measured in the $j^{\rm th}$ stratified bin in $(W_{5891},z_{\rm abs})$. In each such bin, $P_{\rm visual}$ is computed as the mean visual score of the inspected candidates.

Because the visual-inspection sample is stratified in equivalent width and redshift, the quoted global purity is computed as the catalog-weighted mean of the binned visual-purity values assigned to the final absorber catalog, not as the unweighted mean over the inspected systems. This gives a catalog-weighted visual purity of \(\simeq77\%\) for the full catalog and a similar value for the fiducial statistical sample. The two-dimensional visual-purity map is discussed in Appendix~\ref{appendix:selection_function}.

\begin{figure}
    \centering
    \includegraphics[width=\columnwidth]
    {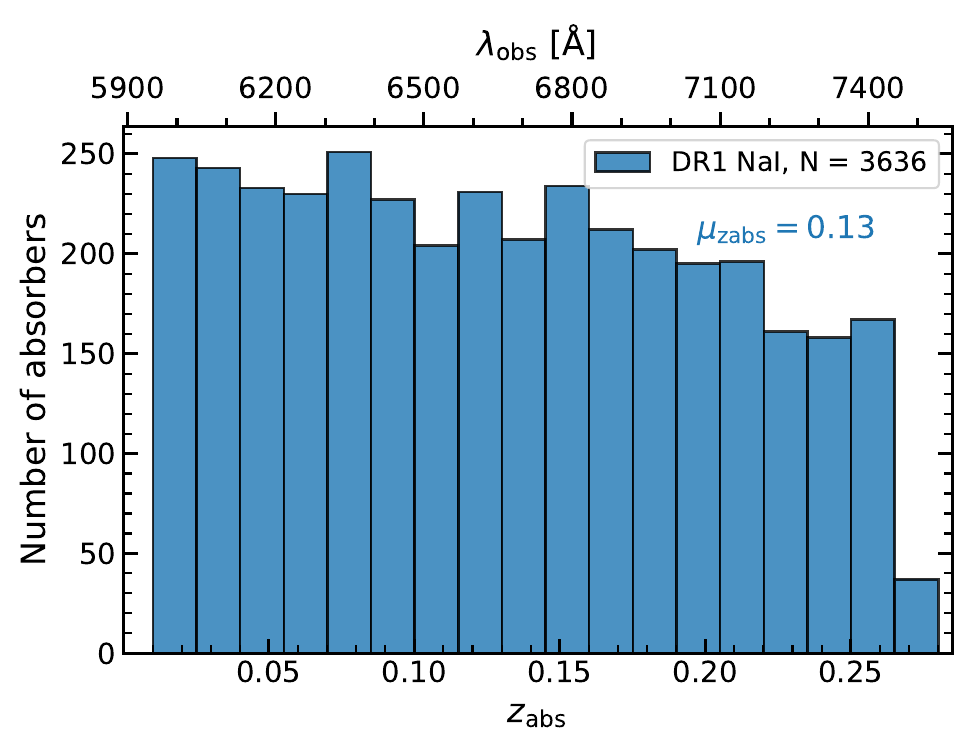}
    \caption{Observed redshift distribution of the 3636 \nai\ absorbers in the final DESI DR1 catalog. The mean absorber redshift is $\langle z_{\rm abs}\rangle=0.13$.}
    \label{fig:redshift}
\end{figure}

\section{\nai~Absorber Catalog}
\label{results}

In this section, we first examine the observed redshift,
equivalent-width, doublet-ratio, and column-density distributions of our \nai\ absorber catalog. Second, we describe the adopted statistical sample and the purity and completeness correction weights applied to it. Finally, we measure the evolution of the absorber incidence and cosmic mass density as functions of redshift and rest-frame equivalent width.

\begin{figure*}
    \centering
    \includegraphics[width=0.42\linewidth]
    {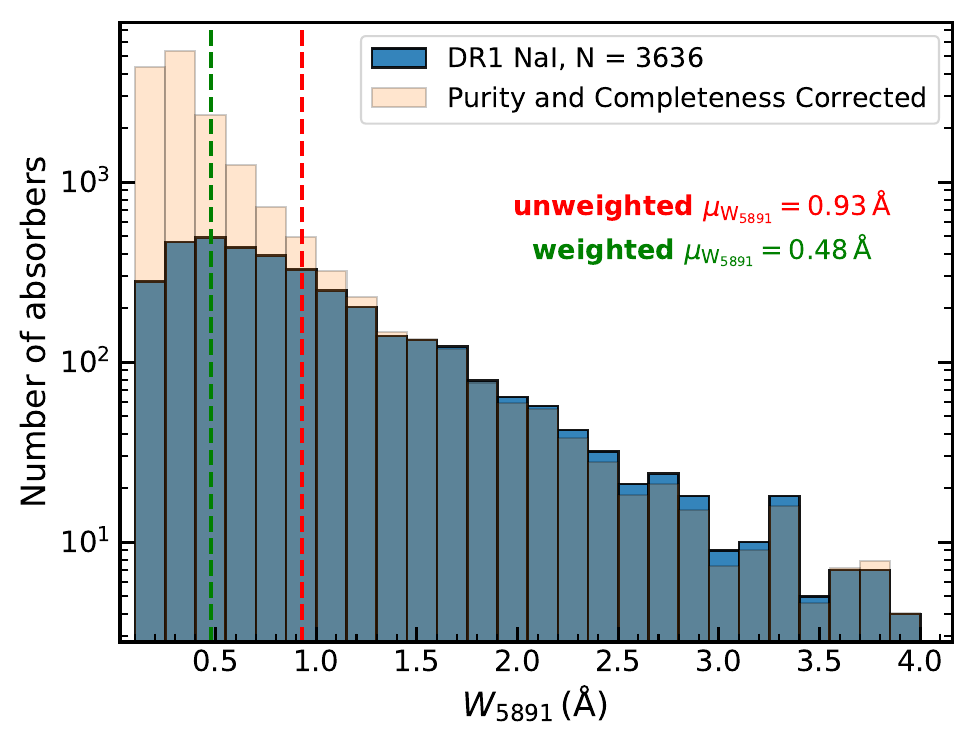}
    \includegraphics[width=0.43\linewidth]
    {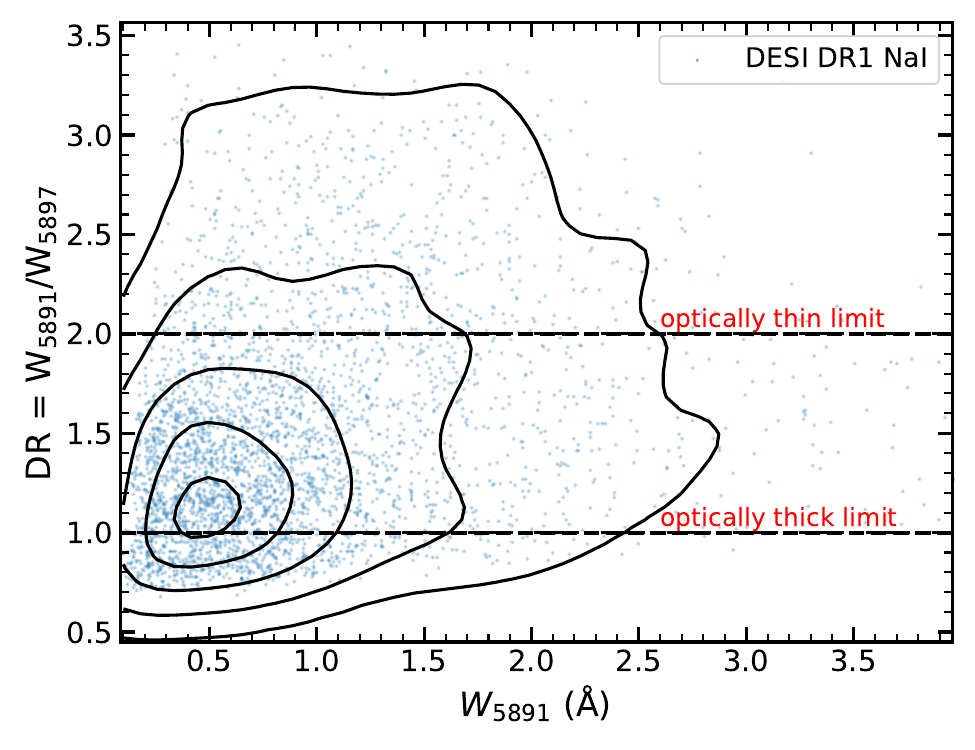}\\
    \includegraphics[width=0.43\linewidth]
    {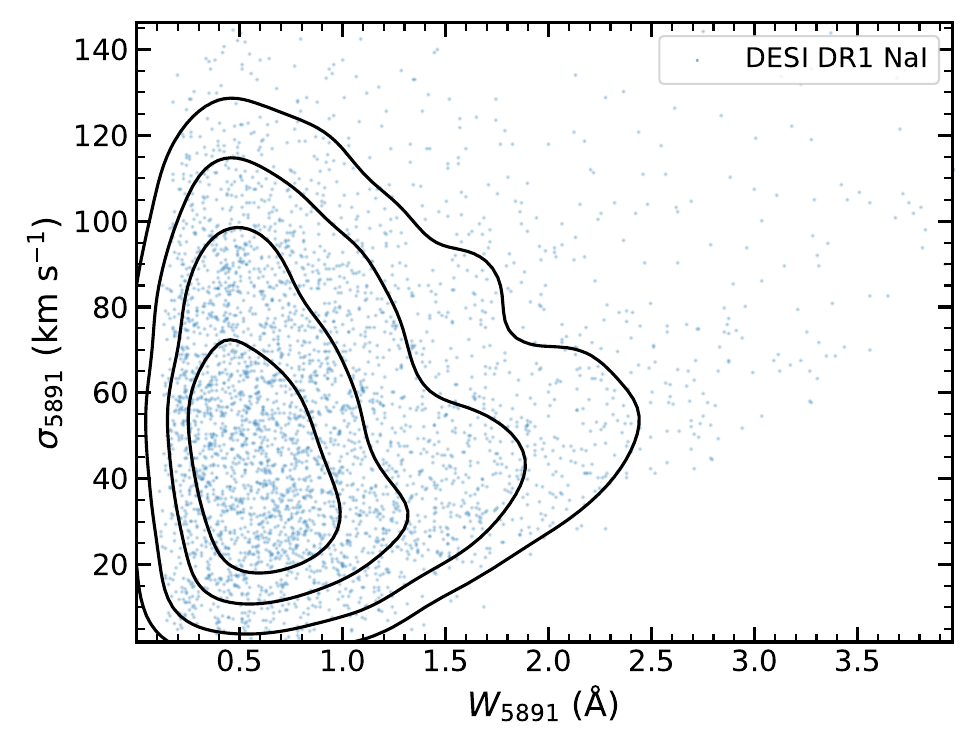}
    \includegraphics[width=0.45\linewidth]
    {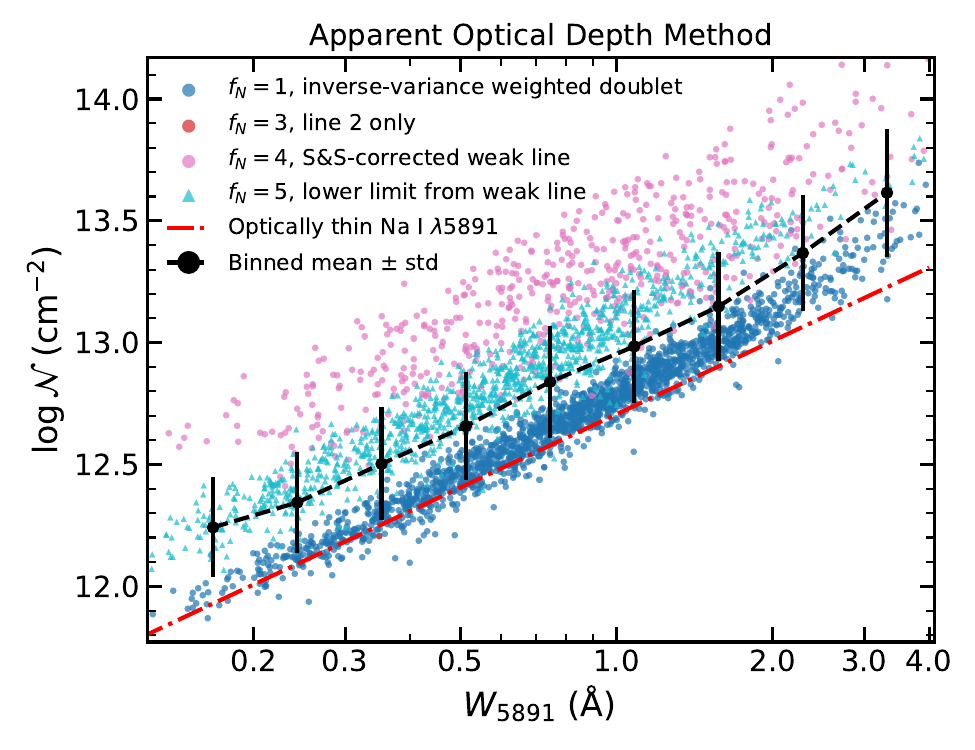}
    \caption{Properties of the individual \nai\ absorbers. \textbf{Upper left}: The observed and purity and completeness-corrected $W_{5891}$ distributions. \textbf{Upper right}: doublet ratio as a function of $W_{5891}$. The dashed
    lines mark the optically thick ($DR=1$) and optically thin
    ($DR=2$) limits. \textbf{Lower left}: corrected velocity width of the $\lambda5891$ line, $\sigma_{5891}$, as a function of $W_{5891}$. \textbf{Lower right}: AODM \nai\ column density as a function of $W_{5891}$. The different  markers indicate the column-density measurement flags described in Section~\ref{column_density}. The black points and error bars show the mean and standard deviation of the column-density distribution. The dotted-dashed curve shows the column density expected in the optically thin limit from the curve of growth. Only systems with finite AODM column-density estimate ($f_N\neq7$) are included in this panel. The $f_N=2$ and $f_N=6$ classes are absent because no systems have these classifications. For upper right and lower left, the black contours mark the 5th, 25th, 50th, 75th, and 95th percentiles.}
    \label{fig:nai_properties}
\end{figure*}

\subsection{Properties of Individual Systems}
\label{properties}

The final catalog contains $3636$ \nai\ absorbers over
$0.01\lesssim z_{\rm abs}\lesssim0.27$, with a mean redshift of $\langle z_{\rm abs}\rangle=0.13$. The redshift distribution is shown in Figure~\ref{fig:redshift}.

Figure~\ref{fig:nai_properties} summarizes the measured properties of the individual \nai\ absorbers. The equivalent-width distribution (top left panel) is strongly weighted toward weak systems and declines steadily toward larger $W_{5891}$, with absorbers above $W_{5891}\simeq2\,\Ang$ becoming relatively uncommon. The purity and completeness corrections have their largest effect at low equivalent width, where the recovery completeness is lowest, and further emphasizes that weak absorbers dominate the population over the measured range.

The doublet ratio as a function of $W_{5891}$ is shown in the top-right
panel. Most systems lie between the optically thick (${\rm DR}=1$) and
optically thin (${\rm DR}=2$) limits, with a large fraction closer to the
saturated limit, indicating that saturation is common in the catalog.
The scatter outside these limits is primarily associated with measurement
uncertainties, particularly in the weaker $\lambda5897$ component. 

The bottom-left panel shows the instrumental-resolution-corrected velocity
width, $\sigma_{5891}$, as a function of $W_{5891}$. Stronger absorbers
generally have broader profiles, showing that the increase in equivalent
width is accompanied by an increasing velocity extent of the absorption.
Thus, the doublet ratio primarily provides information on saturation, while
$\sigma_{5891}$ provides complementary information on the kinematic width
of the absorbing gas.

The inferred \nai\ column density increases with equivalent width (bottom right panel), as expected. For systems with finite AODM measurements, the column densities span approximately $12.05\lesssim\log\mathcal{N}(\mathrm{Na\,I})/\mathrm{cm}^{-2}]\lesssim13.8$. Column-density flags distinguish weighted doublet measurements, single-line measurements, saturation-corrected estimates, and lower limits. Most systems lie above the linear, optically thin part of the curve of growth (dot-dashed line), indicating that the \nai\ absorption is generally saturated in our catalog.

We also construct median absorber-frame composite spectra in bins of
$W_{5891}$. The \nai\ doublet is clearly detected in all bins, while no
robust \caii\ H\&K absorption is detected. The stacking procedure and
quantitative equivalent-width measurements are presented in
Appendix~\ref{appendix:residual_stacks}. 

\subsection{Absorber Incidence and Equivalent-width Distribution}\label{incidence_rates}

We measure the cosmic incidence of \nai~absorbers following the standard absorber-statistics formalism used in previous surveys \citep{nestor05,cooksey10,zhu13a,abbas2024,anand2025}. We restrict the statistical analysis to systems with
$0.95 < W_{5891} < 3.5\,\Ang$, where the catalog corrections are
well constrained as the completeness is above $50\%$. This defines our \textit{fiducial sample}, consisting of $1336$ \nai\ absorbers. For all statistical measurements, we use the purity- and completeness-corrected weights defined in Eqn.~\ref{eq:nai_weight}. These weights correct the observed absorber counts for detection incompleteness and residual contamination. As a robustness check, we also repeat the measurements using completeness-only weights, \(w_i=1/C_i\) as described in Section~\ref{sec:robustness}. We define the redshift path sensitivity function \citep{lanzetta1987,steidel1992} as
\begin{equation}
    g(z)
    =
    \sum_{q=1}^{N_{\rm qso}}
    \sum_{k=1}^{N_{{\rm int},q}}
    \mathbb{I}
    \left(
    z_{{\rm min},qk}
    \leq z <
    z_{{\rm max},qk}
    \right),
    \label{eq:redshift_sensitivity}
\end{equation}
where $N_{\rm qso}$ is the number of QSO sightlines in the final parent sample, $N_{{\rm int},q}$ is the number of usable redshift intervals along the $q$th sightline, and $z_{{\rm min},qk}$ and $z_{{\rm max},qk}$ are, respectively, the lower and upper absorber-redshift boundaries of the $k$th usable interval along that sightline. The quantity $\mathbb{I}$ is the indicator function. Thus, $g(z)$ gives the number of sightlines over which the \nai\ doublet could be searched at absorber redshift $z$. The usable intervals account for the wavelength boundaries and all masks applied during the absorber search.

For each redshift interval $[z_1,z_2]$, we compute the searchable
absorption-distance path. For the flat Universe cosmology, we define
\begin{equation}
    E(z)
    \equiv
    \frac{H(z)}{H_0}
    =
    \left[
    \Omega_{\rm m}(1+z)^3 + \Omega_{\Lambda}
    \right]^{1/2}.
    \label{eq:ez}
\end{equation}
The relevant path elements \citep{bahcall1969} are then
\begin{equation}
\begin{aligned}
    \frac{dX}{dz}
    &=
    \frac{(1+z)^2}{E(z)} 
\end{aligned}
\label{eq:path_elements}
\end{equation}

The total searchable paths are
\begin{equation}
\begin{aligned}
    \Delta X
    &=
    \int_{z_1}^{z_2}
    g(z)\frac{dX}{dz}\,dz
\end{aligned}
\label{eq:survey_paths}
\end{equation}

Here, $\Delta X$ represents the total absorption-distance path sampled by the survey. The corrected incidence per unit absorption survey path is 
\begin{equation}
    \frac{dN}{dX}
    =
    \frac{1}{\Delta X}
    \sum_i w_i
    \label{eq:dndx_dndt}
\end{equation}
where the summation includes all absorbers within the redshift interval and $w_i=P_i/C_i$ is the combined visual-purity and completeness correction.

The quantity $dN/dX$ is preferred over $dN/dz$ because it removes the cosmological path-length dependence and is directly proportional to the product of the comoving number density of the absorber-hosting structures ($n_{\rm com}$)
and their effective \nai\ absorption cross-section ($\sigma_{\rm eff}$).

\begin{figure*}
    \centering
    \includegraphics[width=0.45\linewidth]{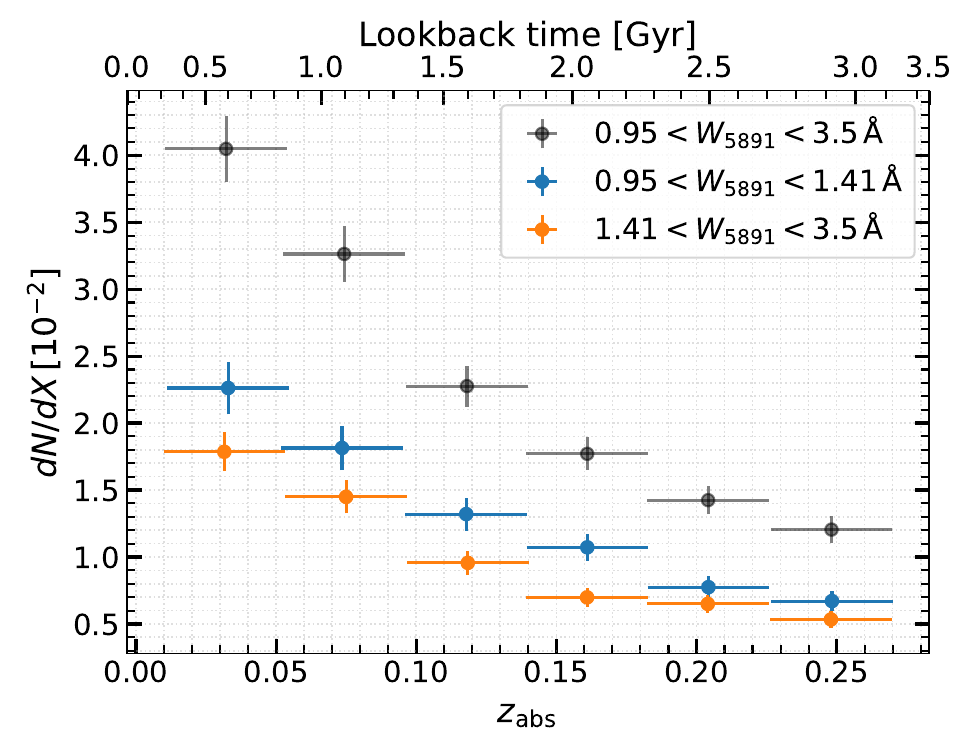}
    \includegraphics[width=0.425\linewidth]{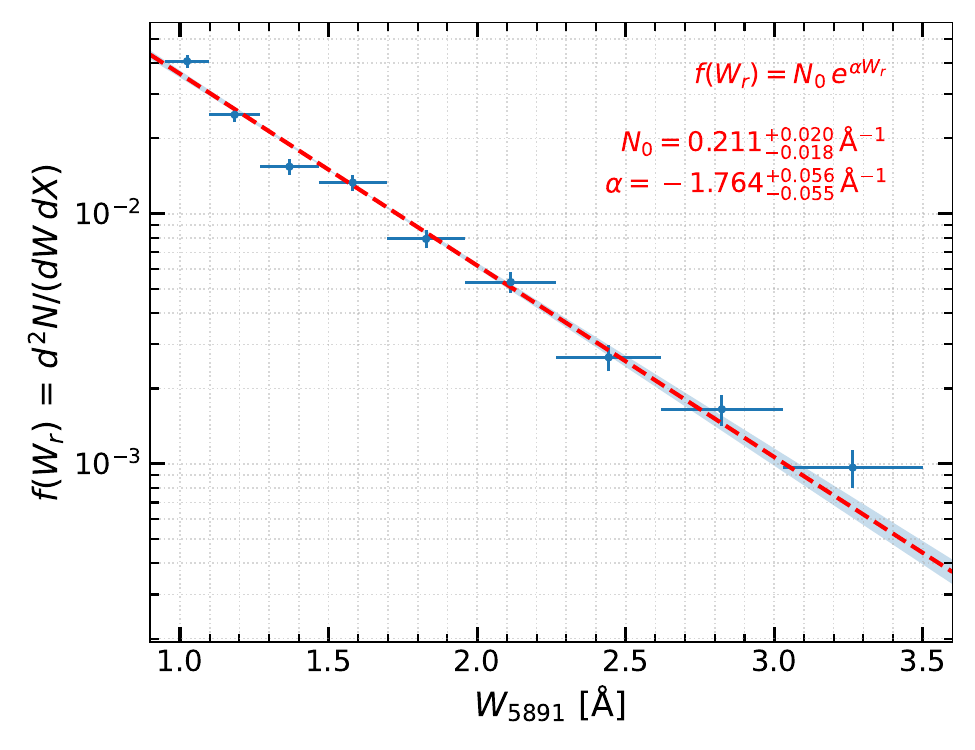}
    \caption{Purity- and completeness-corrected incidence of the intervening \nai\ absorber population as a function of absorber redshift for the fiducial sample with $0.95 < W_{5891} < 3.5\,\Ang$. \textbf{Left:} Incidence per unit absorption path, $dN/dX$. The sample is divided at its median equivalent width,
    $W_{5891}=1.41\,\Ang$, into lower-EW ($0.95<W_{5891}<1.41\,\Ang$) and higher-EW ($1.41<W_{5891}<3.5\,\Ang$) subsamples.
    Vertical error bars show the weighted-counting uncertainties propagated to $dN/dX$, while horizontal error bars indicate the
    redshift-bin widths. The upper axis shows the corresponding lookback time. \textbf{Right:} The rest-frame equivalent-width distribution, $f(W_r)=d^2N/(dW_r\,dX)$, for the same fiducial statistical sample. Vertical error bars show the weighted-counting uncertainties propagated to $f(W_r)$, while horizontal error bars indicate the equivalent-width bin half-widths. The red dashed curve shows the best-fitting exponential model, and the shaded region marks the 16th--84th percentile range obtained from bootstrap resampling.}
    \label{fig:nai_incidence}
\end{figure*}

We also measure the differential rest-frame equivalent-width frequency distribution of the stronger \ion{Na}{1}~$\lambda5891$ transition,
\begin{equation}
    f(W_r)
    \equiv
    \frac{d^2N}{dW_r\,dX}
    =
    \frac{1}{\Delta W_r\,\Delta X}
    \sum_i w_i ,
    \label{eq:fw_eqn}
\end{equation}
where the sum is evaluated over absorbers in an equivalent-width bin of width $\Delta W_r$, and we denote $W_r\equiv W_{5891}$. We measure $f(W_r)$ over the fiducial equivalent-width range
$0.95 < W_r < 3.5\,\Ang$. The weighted-counting uncertainty in each bin is
\begin{equation}
    \sigma_{\rm count}
    =
    \left(
    \sum_i w_i^2
    \right)^{1/2},
    \label{eq:weighted_count_error}
\end{equation}
where the sum is over absorbers in the corresponding redshift or
equivalent-width bin. The uncertainties on the incidence and
equivalent-width frequency measurements are then
\begin{equation}
\begin{aligned}
    \sigma_{dN/dX}
    =
    \frac{\sigma_{\rm count}}{\Delta X},\,\,
    \sigma_{f(W_r)}
    &=
    \frac{\sigma_{\rm count}}
    {\Delta W_r\,\Delta X} 
\end{aligned}
\label{eq:incidence_errors}
\end{equation}
These uncertainties represent weighted-counting errors propagated through the corresponding absorption-path. They do
not include the finite-sampling uncertainties in the binned visual-purity estimates, which should be regarded as an additional systematic uncertainty in the purity correction. The statistical uncertainties in the completeness estimates are much smaller and are not included in the quoted error bars.

The left panel of Figure~\ref{fig:nai_incidence} shows the purity- and
completeness-corrected incidence for the full fiducial sample and after
dividing the sample at its median equivalent width,
$W_{5891}=1.41\,\Ang$. This separates the absorbers into lower-EW
($0.95<W_{5891}<1.41\,\Ang$) and higher-EW
($1.41<W_{5891}<3.5\,\Ang$) subsamples containing equal numbers of
systems. This splitting is not intended to represent physically distinct
absorber populations.

The full fiducial sample shows a clear decrease in $dN/dX$ with increasing
redshift, from $\simeq(4.05\pm0.24)\times10^{-2}$ at
$z_{\rm abs}\simeq0.03$ to $\simeq(1.20\pm0.10)\times10^{-2}$ at
$z_{\rm abs}\simeq0.25$, corresponding to an increase by a factor of
$\simeq3.4\pm 0.34$ toward the present epoch. Both equivalent-width subsamples show
the same trend. For lower-EW systems
($0.95<W_{5891}<1.41\,\Ang$), $dN/dX$ decreases from
$\simeq(2.3\pm0.2)\times10^{-2}$ at $z_{\rm abs}\simeq0.03$ to
$\simeq(0.7\pm0.08)\times10^{-2}$ at $z_{\rm abs}\simeq0.25$,
corresponding to an increase by a factor of $\simeq3.4\pm0.5$ toward the
present epoch. The higher-EW systems ($1.41<W_{5891}<3.5\,\Ang$) show a
similar decrease, from $dN/dX\simeq(1.8\pm0.14)\times10^{-2}$ at
$z_{\rm abs}\simeq0.03$ to $dN/dX\simeq(0.5\pm0.06)\times10^{-2}$ at
$z_{\rm abs}\simeq0.25$. The rise toward low redshift is therefore present in the full sample and in
both equivalent-width subsamples. The upper axis shows the corresponding
lookback time, providing a direct view of the temporal evolution. The
measurements suggest that the rise in $dN/dX$ becomes more pronounced over
the last $\sim2$~Gyr.

The cosmic absorption-path incidence can also be expressed in terms of the
physical quantity that it traces \citep{bahcall1969},
\begin{equation}
    \frac{dN}{dX} = \frac{c}{H_0}n_{\rm com}\sigma_{\rm eff}.
    \label{eq:nsigma}
\end{equation}
The observed increase in $dN/dX$ implies a corresponding increase in
$n_{\rm com}\sigma_{\rm eff}$ toward the present epoch. Thus, the evolution
reflects an increase in the abundance and/or effective cross-section of the
\nai-bearing structures.

In the right panel of Figure~\ref{fig:nai_incidence}, we show the rest-frame equivalent-width frequency distribution, $f(W_r)$. The distribution declines smoothly with increasing equivalent width, demonstrating that strong \nai~absorbers are less common than weaker systems. Following the parameterization commonly adopted for intervening metal-absorber populations
\citep[e.g.,][]{nestor05,cooksey10,anand2025}, we model the
distribution as
\begin{equation}
    f(W_r)=N_0\exp(\alpha W_r)
    \label{eq:exponential}
\end{equation}
We determine the exponential parameters using an unbinned
maximum-likelihood fit over $0.95 < W_{5891} < 3.5 \,\Ang$, accounting for the purity and completeness weights of each absorber and the finite fitted equivalent-width interval. Uncertainties are estimated by bootstrap resampling the absorber sample, with the quoted limits corresponding to the 16th and 84th percentiles of the resulting parameter distributions. The best-fitting parameters and their uncertainties are
\begin{equation}
    N_0
    =
    0.211^{+0.020}_{-0.018}\,
    \Ang^{-1},
    \qquad
    \alpha
    =
    -1.764^{+0.056}_{-0.055}\,
    \Ang^{-1}
\end{equation}
The exponential model describes the measured distribution fairly well, although the highest-equivalent-width bin lies modestly above the model prediction. 

To test for evolution in the shape of the equivalent-width distribution, we repeat the fit to $f(W_r)$ in four redshift bins defined by the quartiles of the fiducial absorber sample. These bins are listed in Table~\ref{tab:parameters}. The fitted slopes are consistent within their uncertainties, with
$W_\ast=-1/\alpha$ remaining near $0.55$--$0.60\,\Ang$ in all redshift bins, while the normalization decreases monotonically with increasing redshift, consistent with the evolution in $dN/dX$. For the full sample, $W_\ast=0.567^{+0.019}_{-0.017}\,\Ang$. Integrating the fitted distributions over \(0.95 < W_{5891} < 3.5 \,\Ang\) gives values consistent with the directly measured \(dN/dX\) in all four redshift intervals. This comparison is not independent of the weighted counts used in the fit, but shows that the exponential form adequately describes the equivalent-width distribution over the fitted range. To our knowledge, this represents the first statistical measurement of the cosmic incidence of intervening \nai\ absorbers at these epochs.

\begin{deluxetable}{cccccccccc}
    \tablecaption{
    Redshift evolution of the \nai\ equivalent-width distribution
    (Eqn.~\ref{eq:exponential}) for our fiducial
    $0.95 < W_{5891} < 3.5 \,\Ang$ statistical sample.
    \label{tab:parameters}
    }
    \tablehead{
    \colhead{$z_{\rm abs}$} &
    \colhead{$\langle z \rangle$} &
    \colhead{$N_{\rm abs}$} &
    \colhead{$N_{\rm corr}$} &
    \colhead{$\Delta X$} &
    \colhead{$(dN/dX)_{\rm obs}$} &
    \colhead{$N_0$} &
    \colhead{$\alpha$} &
    \colhead{$W_\ast=-1/\alpha$} &
    \colhead{$(dN/dX)_{\rm exp}$} \\
    &
    &
    &
    &
    &
    &
    \colhead{($\Ang^{-1}$)} &
    \colhead{($\Ang^{-1}$)} &
    \colhead{($\Ang$)} &
    }
    \startdata
    $0.010$--$0.060$ & 0.036 & 336 & 428.6& 10772.3 &
    $0.040\pm0.002$ &
    $0.395^{+0.074}_{-0.060}$ &
    $-1.792^{+0.104}_{-0.111}$ &
    $0.558^{+0.034}_{-0.032}$ &
    $0.040\pm0.009$ \\
    $0.060$--$0.118$ & 0.087 & 328 & 386.8& 13419.9 &
    $0.029\pm0.002$ &
    $0.283^{+0.053}_{-0.046}$ &
    $-1.783^{+0.107}_{-0.111}$ &
    $0.561^{+0.036}_{-0.033}$ &
    $0.029\pm0.007$ \\
    $0.118$--$0.185$ & 0.151 & 335 & 330.5& 17604.3 &
    $0.019\pm0.001$ &
    $0.193^{+0.038}_{-0.028}$ &
    $-1.816^{+0.104}_{-0.120}$ &
    $0.551^{+0.034}_{-0.034}$ &
    $0.019\pm0.004$ \\
    $0.185$--$0.270$ & 0.224 & 337 & 320.2& 24301.5 &
    $0.013\pm0.001$ &
    $0.105^{+0.021}_{-0.016}$ &
    $-1.647^{+0.111}_{-0.122}$ &
    $0.607^{+0.044}_{-0.042}$ &
    $0.013\pm0.003$ \\
    $0.010$--$0.270$ & 0.13 & 1336 & 1466.1& 66098.0 &
    $0.022\pm0.001$ &
    $0.211^{+0.020}_{-0.018}$ &
    $-1.764^{+0.056}_{-0.055}$ &
    $0.567^{+0.019}_{-0.017}$ &
    $0.022\pm0.003$
    \enddata
    
    \tablecomments{
    The equivalent-width distributions are fitted with
    $f(W_r)=N_0\exp(\alpha W_r)$, where $W_r\equiv W_{5891}$.
    $N_{\rm abs}$ is the unweighted absorber count, $N_{\rm corr}$ is the purity- and completeness-corrected absorber count, and $\Delta X$ is the total absorption path. Because each absorber is weighted by the ratio of purity to completeness, $N_{\rm corr}$ can be smaller than $N_{\rm abs}$ when the purity correction outweighs the completeness correction, as in the highest-redshift bins. $(dN/dX)_{\rm obs}$ is measured directly from the purity- and completeness-corrected counts, whereas $(dN/dX)_{\rm exp}$ is obtained by integrating the fitted exponential model over the same equivalent-width interval. Quoted fit uncertainties correspond to the 16th and 84th percentiles of the bootstrap distributions. The last row gives the result for the full redshift interval.}
\end{deluxetable}

\subsection{Cosmic Mass Density of \nai\ Absorbers}\label{sec:omega_nai}

The cosmic mass density of an ionic species gives a column-density-weighted measure of its abundance relative to the present-day critical density of the Universe. We compute the
purity- and completeness-corrected lower limits on cosmic mass density of \nai\ absorbers with $0.95<W_{5891}<3.5\,\Ang$. In $\Omega_{\rm NaI}$ calculation, we  only include systems with finite AODM column-density estimates ($f_N\leq 4$), yielding a total of $1013$ systems ($\simeq76\%$ of the fiducial sample). Systems with lower-limit column-density ($f_N = 5,6$) estimates are excluded from the fiducial measurement and considered separately as a robustness check. In each redshift bin,
\begin{equation}
    \Omega_{\rm Na\,I}(z)
    =
    \frac{H_0m_{\rm Na}}
         {c\rho_{\rm crit,0}}
    \frac{
    \displaystyle\sum_i
    w_i\,\mathcal{N}_i({\rm Na\,I})
    }
    {\Delta X(z)} ,
    \label{eq:omega_nai}
\end{equation}
where $\mathcal{N}_i({\rm Na\,I})$ is the \nai\ column density of the $i$th absorber, $\Delta X$ is the absorption path, and $w_i=P_i/C_i$ is the combined purity and completeness weight. $H_0$ is the present-day Hubble constant, $m_{\rm Na}$ is the mass of a sodium atom, $c$ is the speed of light, and $\rho_{\rm crit,0}$ is the present-day critical density, which provides a fixed reference density for comparing $\Omega_{\rm Na\,I}$ across redshift. The uncertainty is estimated by propagating both the column-density errors and the weighted counting uncertainty,
\begin{equation}
    \sigma_{\Omega_{\rm Na\,I}}
    =
    \Omega_{\rm Na\,I}
    \left[
    \frac{\sum_i w_i^2\sigma_{\mathcal{N}_i}^2}
    {\left(\sum_i w_i\mathcal{N}_i\right)^2}
    +
    \frac{\sum_i w_i^2}
    {\left(\sum_i w_i\right)^2}
    \right]^{1/2},
\end{equation}
where $\sigma_{\mathcal{N}_i}$ is the uncertainty on the column density.

\begin{figure*}
    \centering
    \includegraphics[width=0.45\linewidth]{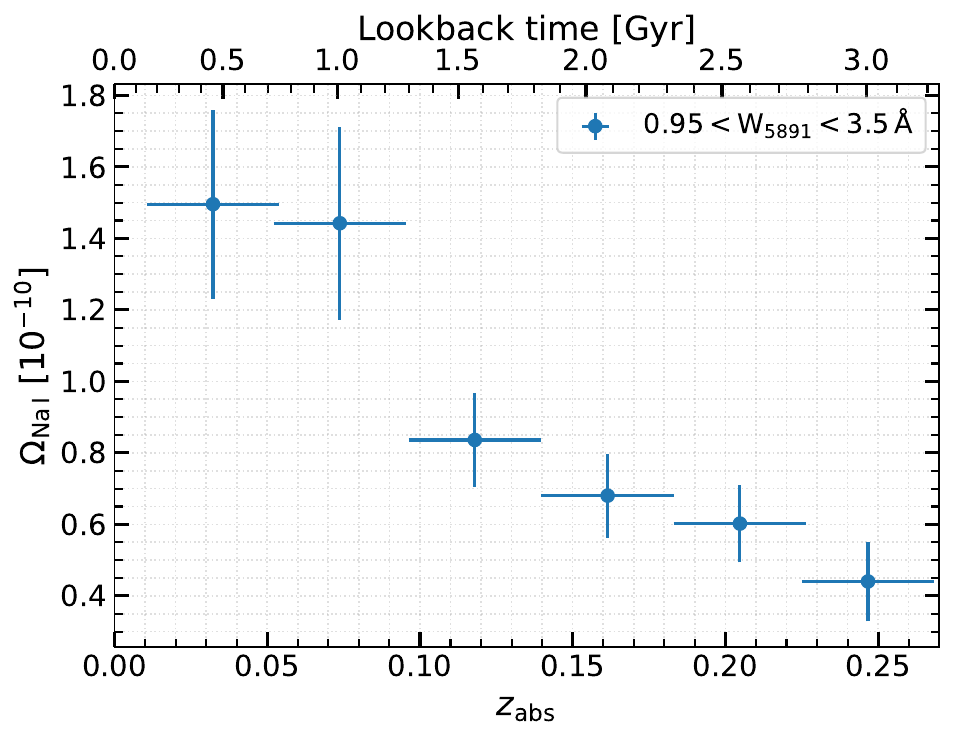}
    \includegraphics[width=0.45\linewidth]{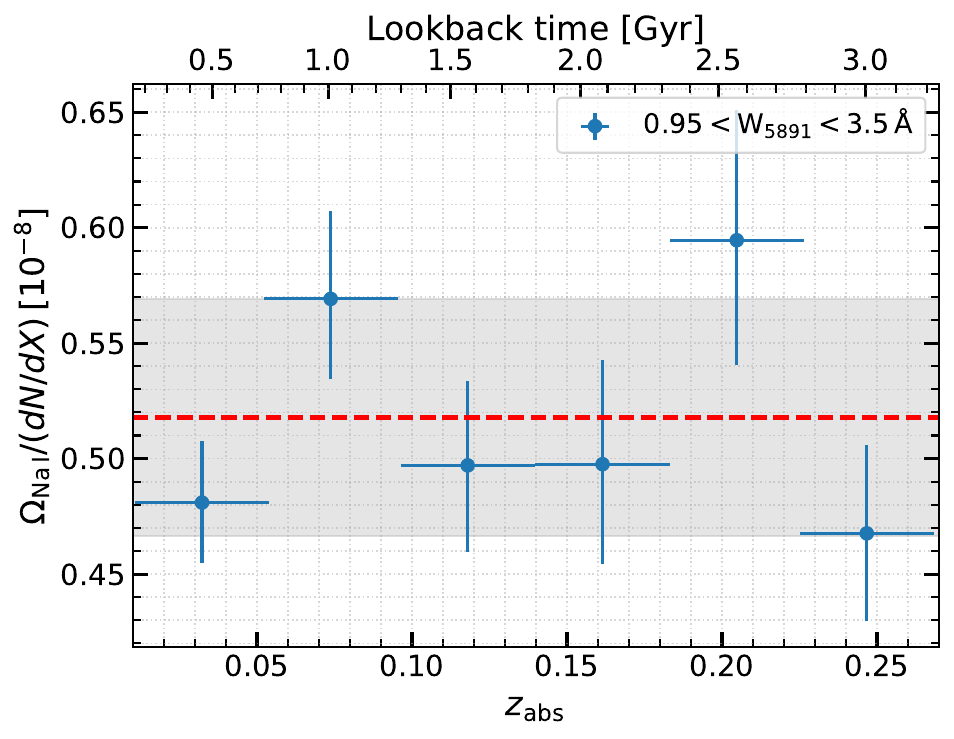}
    \caption{\textbf{Left:} Purity- and completeness-corrected cosmic mass density of neutral sodium, $\Omega_{\rm Na\,I}$, as a function of absorber redshift. \textbf{Right:} Ratio $\Omega_{\rm Na\,I}/(dN/dX)$ as a function of absorber redshift, which is proportional to the purity- and completeness-weighted mean measured \nai\ column density per absorber. The horizontal dashed line marks the mean of the binned measurements, while the horizontal shaded band indicates their $1\sigma$ standard deviation. The upper axes give the corresponding lookback time calculated using the adopted Planck cosmology. In the left panel, vertical error bars include the column-density and weighted-counting
    uncertainties. In the right panel, vertical error bars are estimated by bootstrap resampling the absorbers within each redshift bin. Horizontal error bars indicate the redshift-bin widths.}
    \label{fig:nai_omega}
\end{figure*}

The left panel of Figure~\ref{fig:nai_omega} shows that
$\Omega_{\rm Na\,I}$ increases toward the present epoch, from
$\simeq(0.44\pm0.11)\times10^{-10}$ at $z_{\rm abs}\simeq0.25$ to
$\simeq(1.50\pm0.26)\times10^{-10}$ at $z_{\rm abs}\simeq0.03$. This corresponds to an increase by a factor of $\simeq3.4\pm1.05$ over the last $\simeq2.5$~Gyr, and closely follows the evolution seen in the absorber incidence.

To separate changes in incidence from changes in typical absorber column density, we also examine the ratio
\begin{equation}
    \frac{\Omega_{\rm Na\,I}}{dN/dX}
    =
    \frac{H_0m_{\rm Na}}
         {c\rho_{\rm crit,0}}
    \left\langle
    \mathcal{N}({\rm Na\,I})
    \right\rangle_w ,
    \end{equation}
    where
    \begin{equation}
    \left\langle
    \mathcal{N}({\rm Na\,I})
    \right\rangle_w
    =
    \frac{
    \sum_i w_i\mathcal{N}_i({\rm Na\,I})
    }{
    \sum_i w_i
    }
    \label{eqn:omega_nai}
\end{equation}
is the purity- and completeness-weighted mean \nai\ column density per absorber. The error bars on this ratio are estimated by bootstrapping the absorbers within each redshift bin. For this ratio, we recompute $dN/dX$ using the same subset of absorbers with finite AODM column-density measurements used to calculate $\Omega_{\rm Na\,I}$. This ensures a consistent comparison, with the numerator and denominator computed from the same absorber population. The right panel of Figure~\ref{fig:nai_omega} shows that $\Omega_{\rm Na\,I}/(dN/dX)$ has no significant redshift evolution, indicating little change in the mean measured \nai\ column density per absorber.


\subsection{Robustness of Observed Trends}\label{sec:robustness}

We performed several checks to test whether the observed redshift evolution depends on the sample definition, correction scheme, or binning choice. Restricting the sample to $z_{\rm abs}>0.05$, or adopting a more conservative
equivalent-width threshold of $W_{5891}>1.5\,\Ang$, leaves the main trends unchanged. In both cases, the incidence continues to increase toward lower redshift, while the equivalent-width distribution and cosmic-density evolution remain consistent with the fiducial results.

We also repeated the measurements using alternative statistical weights. The fiducial analysis uses $w_i=P_i/C_i$, which corrects for both visual purity and completeness. We repeated the calculation using completeness-only weights, $w_i=1/C_i$, and also using unweighted counts, $w_i=1$. These choices change the normalization, but the qualitative redshift trends in $dN/dX$, $f(W_r)$, and $\Omega_{\rm Na\,I}$ remain the same. Therefore, the observed evolution is not introduced by the visual-purity correction or by a small number of highly
weighted systems. Furthermore, including systems with lower-limit column-density estimates raises the normalization of $\Omega_{\rm Na\,I}$ but preserves the same qualitative redshift evolution.

Finally, we tested alternative redshift binning schemes, including broader fixed-redshift bins and equal-count bins. The overall evolution of $dN/dX$, $\Omega_{\rm Na\,I}$, and their ratio remains consistent, with only the expected changes in bin-to-bin fluctuations. These tests indicate that the main trends are robust and not driven by the lowest-redshift absorbers, the adopted purity/completeness weights, or one particular choice of redshift bins. Results for alternative redshift binning and statistical weighting choices are shown in Appendix~\ref{appendix:robustness}.
\section{Discussion}\label{discussion}

\subsection{Physical Nature of the Intervening \nai\ Absorbers}
\label{nai_nature}

Neutral sodium has a low first ionization potential, $5.14~{\rm eV}$, so
detectable \nai\ absorption preferentially arises in dense, well-shielded
regions of atomic gas \citep[see][for recent reviews]{churchill2025a,churchill2025b}.
\nai\ therefore traces only a restricted component of the neutral-gas
reservoir. Consequently, several empirical relations have been established
between $\mathcal{N}(\mathrm{Na\,I})$ and $\mathcal{N}(\mathrm{H\,I})$ in
the Milky Way and external galaxies, albeit with substantial scatter
\citep{ferlet1985,sembach1993,moretti2025}.

The $\mathcal{N}(\mathrm{Na\,I})$ values inferred from the AODM place these
absorbers in the context of established \nai--\hi\ relations. For example,
the classical Galactic relation \citep{sembach1993}, valid over the range
$\rm 11.6\leq \log\mathcal{N}(\mathrm{Na\,I}) \leq 13.4$, gives
\begin{equation}
    \log \mathcal{N}(\mathrm{H\,I}) =
    0.688\log \mathcal{N}(\mathrm{Na\,I}) + 12.16.
\end{equation}
For the range of \nai\ column densities measured in our sample, this
relation gives
$\log\,[\mathcal{N}(\mathrm{H\,I})/{\rm cm}^{-2}]\simeq20.4$--$21.7$.
Here, we simply extrapolate this relation to our highest
$\mathcal{N}(\mathrm{Na\,I})$ values. These values place many of the
stronger systems in the DLA regime and support the view that strong
intervening \nai\ absorption traces high-column-density neutral gas.

Recent JWST work has revisited this conversion in a high-redshift outflow
system. \citet{moretti2025} emphasize that converting a trace ion such as
\nai\ into \hi\ requires assumptions about the neutral sodium fraction, the
gas-phase abundance, and dust depletion. Using a direct \hi\ measurement in
a $z=2.4$ sub-DLA associated with a massive galaxy outflow, they instead
derive
\begin{equation}
    \log \mathcal{N}(\mathrm{H\,I}) =
    \log \mathcal{N}(\mathrm{Na\,I}) + 7.5.
\end{equation}
This is only $0.14$ dex lower than the commonly used local linear
calibration,
$\log \mathcal{N}(\mathrm{H\,I})=\log \mathcal{N}(\mathrm{Na\,I})+7.64$
\citep{savage1996,rupke2005a,rupke2005b}. Applied to our measured \nai\ columns, this
gives
$\log\,[\mathcal{N}(\mathrm{H\,I})/{\rm cm}^{-2}]\simeq19.6$--$21.3$,
with the median absorber lying near the sub-DLA/DLA boundary. These conversions are only indicative, as the relation has substantial
environmental scatter. It depends on metallicity, the radiation field,
depletion onto dust, and unresolved cloud structure. In addition,
unresolved saturation can make the measured AODM \nai\ column densities
lower limits. We therefore do not convert our $\Omega_{\rm Na\,I}$
measurement into an $\Omega_{\rm H\,I}$ measurement.

\modify{For comparison, estimates of the local DLA incidence inferred from 21-cm measurements show some variation between studies. \citet{zwaan2005} find $dN/dX\simeq0.045\pm0.006$ at $z\simeq0$, while \citet{braun2012} obtain $0.026\pm0.003$. Using UV observations from HST-FOS, \citet{rao2017} followed up \mgii-selected systems, compiled a sample of 70 DLAs, and derived an empirical parametrization of $dN/dz$ over $0<z<5$. Extrapolating this relation to $z=0$ gives a value closer to the lower estimate from \citet{braun2012}. Thus, the local DLA incidence is of order $dN/dX\simeq0.03$--0.05, comparable to the $dN/dX\simeq0.04$ measured for our strong \nai\ absorbers at $z\simeq0.03$. At $z\simeq0.25$, however, the \nai\ incidence decreases to $\simeq0.012$. This comparison is only approximate, since strong \nai\ absorption does not correspond to a unique $N(\mathrm{H\,I})$ threshold. The incidence of sub-DLAs at these redshifts is also less well constrained because direct Ly$\alpha$ surveys require UV spectroscopy and the available samples are small \citep{meiring2011}. The comparison is therefore consistent with strong \nai\ preferentially tracing high-column-density neutral gas, but does not imply a one-to-one correspondence with either the DLA or sub-DLA population.}

As an initial test for dust associated with the \nai-bearing systems, we
compare the broadband colors of quasars hosting \nai\ absorbers with those
of quasars without detected \nai, matched in quasar redshift
($|\Delta z|\leq0.01$) and W1 magnitude
($|\Delta W1|\leq0.1\,\mathrm{mag}$). The absorber sightlines are
systematically redder than the matched controls. The color excess generally
increases with $W_{5891}$, with the strongest trend seen in $r-W1$,
reaching $\simeq0.15$~mag for the highest-equivalent-width systems. The
optical colors, $g-r$ and $r-z$, show smaller excesses of up to
$\simeq0.06$--$0.07$~mag. This preliminary reddening signal is consistent
with dust being associated with the \nai-bearing gas. A more detailed
analysis, including spectral stacking, accounting for other absorption
systems, and extinction-curve modelling, will be presented in future work.

\nai\ absorption has also been observed in a range of
neutral-gas environments, including Galactic disk sightlines, high-velocity
clouds, DLAs, and neutral outflows
\citep{rupke2002,rupke2005a,martin2006,chen2010,richter2011,avery2022}.
Because our survey is blind and does not require a known foreground galaxy,
the catalog likely includes systems arising in more than one of these
environments. The absorber statistics measure the incidence of
\nai-bearing gas, but cannot by themselves identify a unique host
population.

\subsection{Cosmic Evolution of \nai~at Low Redshifts}
\label{nai_evolution}

The absorber statistics allow us to separate incidence evolution from changes
in the typical measured column density. In the standard picture, as shown in
Equation~\ref{eq:nsigma}, $dN/dX \propto n_{\rm com}\,\sigma_{\rm eff}$.
The observed rise in $dN/dX$ therefore implies that the product
$n_{\rm com}\sigma_{\rm eff}$ increases toward the present epoch. The
equivalent-width distribution, incidence, and $\Omega_{\rm Na\,I}$, however,
are not fully independent diagnostics, since they are derived from the same
absorber population and share the same weight corrections. Therefore, their
agreement should be interpreted as an internal consistency check and not as
independent evidence for the evolution.

The equivalent-width distribution shows a consistent picture. In each
redshift bin, it is well described by an exponential form,
$f(W_r)=N_0\exp(\alpha W_r)$. The main evolution is in the normalization,
which decreases with increasing redshift, while the characteristic scale
$W_\ast=-1/\alpha$ changes only weakly (see Table~\ref{tab:parameters}).
This indicates that the evolution is dominated by a change in the overall
abundance of absorbers, with little change in the shape of the
equivalent-width distribution.

A similar picture emerges from $\Omega_{\rm Na\,I}$. As shown in
Equation~\ref{eqn:omega_nai}, $\Omega_{\rm Na\,I}$ depends on both $dN/dX$
and the weighted mean \nai\ column density, so these quantities are not
independent. Their similar redshift evolution, together with the weak
evolution of $\Omega_{\rm Na\,I}/(dN/dX)$, indicates that the increase in
$\Omega_{\rm Na\,I}$ is driven mainly by the higher incidence of
\nai-bearing structures. It also implies that there is no large change in
the mean measured \nai\ column density per absorber.

We note that the cosmic density of neutral gas, $\Omega_{\rm H\,I}$, shows
little evolution over our redshift range
\citep{rhee2013,bera2019,chowdhury2020}. The much stronger evolution of
$\Omega_{\rm Na\,I}$ therefore cannot be attributed simply to an increase
in the total \hi\ reservoir. Moreover, the \nai\ neutral fraction is
sensitive to the radiation field, owing to its ionization potential well
below the \hi\ ionization threshold, as well as to gas density and dust
shielding. The evolution of $\Omega_{\rm Na\,I}$ is therefore governed by a
combination of these physical factors. As discussed above, our analysis
suggests that its evolution is driven primarily by the evolution of the
absorber incidence, $dN/dX$. We therefore focus below on possible physical
drivers of this incidence evolution.

If many systems arise in high-column-density neutral gas associated with
galaxies, the host population becomes important for interpreting the
incidence evolution. If its comoving abundance changes only weakly over
this redshift range, as is broadly consistent with measurements of
star-forming galaxies at $z<1$ \citep{moustakas2013a}, then the observed
rise in $dN/dX$ would point mainly to an increase in the effective
\nai-bearing cross-section or covering fraction.

Under this limiting interpretation, a factor of $\simeq3.4$ increase in
$dN/dX$ would correspond, at fixed host comoving number density, to a
similar increase in $\sigma_{\rm eff}$, or to an increase in the
characteristic absorbing radius by a factor of $\simeq1.8$. However, these
estimates do not measure absorber sizes directly. They only show that the
observed evolution can be described as a growth in the effective area over
which gas is able to produce detectable \nai. Here, $\sigma_{\rm eff}$ is a
\nai-selected area, not necessarily the physical size of an \hi\ disk. In
absorber statistics, changes in incidence can reflect changes in host
abundance, gas cross-section, or covering fraction
\citep[e.g.,][]{nestor05,cooksey10}. For \nai, the relevant area depends on
where the gas is sufficiently dense and shielded for sodium to remain
neutral \citep{richter2011}. Absorber statistics alone cannot separate
these possibilities.

The increasing incidence may also be influenced by the chemical evolution
of the neutral gas. The metallicity and dust-to-gas ratio of neutral
absorbers generally increase toward the present epoch \citep{peroux20a},
which could favor detectable \nai\ through increased sodium abundance and
more effective dust shielding.

Another possible contributor is the declining ionizing background toward
low redshift. Using the \citet{khaire2019} UV-background model, we find that
the \nai\ photoionization rate\footnote{V. Khaire, private communication.} decreases from $\Gamma_{\rm Na\,I}\simeq1.64\times10^{-13}\,{\rm s^{-1}}$ at $z=0.3$ to $5.9\times10^{-14}\,{\rm s^{-1}}$ at $z=0$, a factor of $\simeq2.8$. A weaker ionizing background would favor a larger neutral sodium fraction
and could therefore increase the effective cross-section over which \nai\ absorption is detectable. This change is comparable in magnitude to the
observed factor of $\simeq3.4$ increase in incidence, although the two should not be interpreted as a direct one-to-one correspondence. Because the \nai\ ionization potential is only 5.14~eV, local radiation from the host galaxy can also affect the ionization balance. Over the relatively small redshift interval considered here, such local radiation may mainly
modify the normalization of the \nai\ ionization conditions, but its role cannot be quantified without detailed photoionization modelling.

\subsection{An Exploratory Host-galaxy Check}
\label{sec:nai_hosts}

Next, to test whether the detected NaI absorbers are associated with foreground galaxies, we performed an exploratory cross-match with the DESI spectroscopic galaxy value added catalog \citep{hahn2023BGS,siudek2024,zou2024} at $0.01<z_{\rm gal}<0.27$. The sample contains $\sim4$ million bright galaxies over the redshift range considered here. We required $1<D_{\rm proj}<100~{\rm kpc}$ and $|\Delta v_{\rm abs-gal}|<500~{\rm km\,s^{-1}}$, where $D_{\rm proj}$ is the physical projected separation between absorber and galaxy and $|\Delta v_{\rm abs-gal}|$ is the velocity difference between them. This gives 67 absorber--galaxy pairs, corresponding to 57 unique \nai\ absorbers, with a median impact parameter of $D_{\rm proj}\sim 30\, \rm kpc$. A small number of absorbers have more than one nearby DESI galaxy within the adopted projected-distance and velocity cuts.

As a simple null test, we keep the absorber angular positions fixed and randomly reshuffle the absorber redshifts among the observed sample, while leaving the galaxy catalog unchanged. This preserves the overall observed $n(z_{\rm abs})$ and angular distribution of the absorbers, but does not explicitly enforce the searchable redshift path of each individual QSO
sightline. We therefore use this test only as a qualitative check on the absorber--galaxy association. The randomized catalogs give a median of 14 unique absorber matches, with 16th--84th percentiles of 10--17 and a maximum value of 23. No realization of 100 reached the observed value ($p<0.01$), so the matched systems are well above the chance-association expectation. This supports the conclusion that at least a subset of the \nai\ absorbers is genuinely associated with foreground galaxies.

The matched galaxies include both star-forming and passive systems. Using $\log\,[{\rm sSFR}/yr^{-1}]>-11$ as a simple division \citep{kauffmann2003}, about two thirds are classified as star-forming. This fraction should be interpreted cautiously because the foreground-galaxy sample is not volume complete \citep{hahn2023BGS,desidr1release2025} and the number of matched systems is small. The current sample therefore does not identify a unique host population.

We also estimate the raw detection fraction of \nai\ absorbers around DESI
foreground galaxies. For each projected-separation bin, the denominator is
the number of foreground galaxy--background QSO configurations, while the
numerator is the number containing a fiducial \nai\ absorber,
$0.95<W_{5891}<3.5~\Ang$, satisfying the adopted velocity criterion.
Within $1<D_{\rm proj}<100~{\rm kpc}$, there are $58,552$ such
configurations and $19$ fiducial \nai--galaxy detections. The detection
fraction is highest at small projected separation, reaching $\simeq0.49\%$
for $1<D_{\rm proj}<10~{\rm kpc}$, and falls to $\lesssim0.05\%$ beyond
$D_{\rm proj}\simeq25~{\rm kpc}$.

We emphasize that these raw detection fractions should not be interpreted
as covering fractions, since the varying spectral sensitivity of the
background QSO sightlines is not accounted for. In addition, the DESI
foreground-galaxy sample is incomplete, so this quantity is not a
volume-complete covering fraction. It is instead a simple
galaxy-conditioned diagnostic based on the available foreground
spectroscopy.

For comparison, \citet{nielsen2013} find that low-redshift galaxies
($z<0.359$) have covering fractions of strong \mgii\ absorption,
$W_r(2796)\geq1~\Ang$, of approximately 40\% within 25~kpc, declining to
$\sim15\%$ at 25--50~kpc and only a few percent at 50--100~kpc. Although
these values are not directly comparable to our raw \nai\ detection
fractions, they provide useful context. \mgii\ traces a broader range of
halo gas, while \caii\ and \nai\ preferentially select colder and more
neutral material \citep{nestor2008,zhu2013,sardane2015}.

The low \nai\ detection fraction and its rapid decline with projected
separation suggest that detectable \nai\ has a small effective cross-section
around these galaxies and is preferentially found at relatively small
galactocentric distances. This is consistent with \nai\ tracing compact,
dense, and well-shielded neutral structures associated with the ISM and
inner circumgalactic regions, with little contribution from a widespread
halo-scale phase.

\subsection{Future Observational Tests}
\label{sec:nai_origin}

The host-galaxy check above shows that at least some \ion{Na}{1}~absorbers are associated with foreground galaxies, but it does not yet identify the
dominant physical origin of the population. More complete host
identifications could determine whether these systems are preferentially
associated with galaxy disks, dense halo clouds, recycled gas, group
environments, or neutral outflows. Integral-field spectroscopy would be
particularly useful for this purpose, since it can identify galaxies close
to the quasar sightline and measure their impact parameters, velocity
offsets, star-formation activity, and environment. Quasar sightlines
containing multiple \nai\ absorbers would be especially efficient targets,
allowing the galaxy environments of several absorbers to be investigated
within the same field. Such observations could test whether the observed
redshift evolution is driven mainly by changes in the host population or
by changes in the \nai-bearing gas covering fraction around similar hosts.

\modify{\nai\ is also a well-established tracer of cool neutral gas in galactic outflows. Blueshifted \nai\ D absorption has been observed in star-forming and infrared-luminous galaxies, where it traces dusty neutral material entrained in large-scale winds \citep{rupke2005a,rupke2005b,martin2006,chen2010,avery2022}. In a complementary transverse study of Galaxies on Top of Quasar (GOTOQ) sightlines, \citet{rubin2022} found that most \nai\ components have relatively small velocity offsets from their host galaxies, while their broad profiles require a contribution from extraplanar gas. They interpret these systems primarily in terms of gravitationally bound galactic-fountain material rather than escaping outflows. Some fraction of the intervening systems in our sample may therefore arise in outflowing or fountain-like gas rather than in quiescent disk or inner-CGM structures. Measurements of galaxy orientation and absorber--galaxy velocity offsets from dedicated follow-up observations could help distinguish between these possibilities.}

Associated \caii\ absorption provides an important complementary test.
Since \nai\ preferentially traces denser and more strongly shielded regions
within \caii-bearing gas \citep{nestor2008,sardane2014,sardane2015}, the
redshift evolution of the \caii\ detection fraction and the
$\mathcal{N}(\mathrm{Na\,I})/\mathcal{N}(\mathrm{Ca\,II})$ ratio could help
distinguish changes in cloud incidence from changes in density, depletion,
or ionization. Comparisons with \caii\ and ultraviolet observations of
\mgii\ could further test whether the observed evolution is specific to
the densest neutral-gas phase.

Spectral stacking and extinction-curve modelling of quasars behind \nai\
absorbers, compared with matched control sightlines, can provide stronger
constraints on the associated dust reddening and reveal weaker absorption
features. Direct photoionization modelling of \nai\ could quantify how gas
density, shielding, and the radiation field affect the \nai\ neutral
fraction. Finally, direct \hi\ measurements, where available
\citep{hu2020,chowdhury2020,oyarzun2025}, could test whether the empirical
$\mathcal{N}(\mathrm{Na\,I})$--$\mathcal{N}(\mathrm{H\,I})$ relations calibrated in the Milky
Way and individual systems apply to this cosmologically selected
population.

Together, these measurements can determine whether the observed
low-redshift evolution is driven primarily by changes in the abundance or
covering fraction of dense neutral structures, or by changes in the
physical conditions that allow sodium to remain neutral.

\section{Summary}
\label{summary}

We present, the first blind statistical survey of intervening \nai\ absorbers in quasar spectra. Starting from the DESI DR1 quasar sample, we define a final high-quality search path containing $214,035$ sightlines and identify 3636 \nai\ systems over $0.01\lesssim z_{\rm abs}\lesssim0.27$. This is the largest sample of cosmologically distributed \nai\ absorbers to date. Our main results are as follows:

\begin{itemize}
    \item We quantify catalog completeness with end-to-end injection--recovery simulations and estimate the purity from visual inspection. The catalog is $50\%$ complete at $W_{5891}=0.95\,\Ang$, with a global purity of $\simeq77\%$.
    
    \item The measured \nai\ column densities using AODM span $12.05\lesssim\log[\mathcal{N}(\mathrm{Na\,I})/\mathrm{cm}^{-2}]\lesssim13.8$,
    with a median of $\simeq12.80$. Empirical \nai--\hi\ relations suggest that many systems lie in the sub-DLA/DLA regime.
    
    \item For the fiducial sample, $0.95<W_{5891}<3.5\,\Ang$, the equivalent-width distribution is well described by an exponential function. Its slope remains nearly unchanged with redshift, while the normalization increases toward low redshift.
    
    \item The incidence of \nai\ absorbers increases toward the present epoch in both EW subsamples, split at $W_{5891}=1.41\,\Ang$. From
    $z_{\rm abs}\simeq0.25$ to $z_{\rm abs}\simeq0.03$, $dN/dX$ rises by a
    factor of $\simeq3.4$ in both subsamples, implying substantial evolution in $n_{\rm com}\sigma_{\rm eff}$.
    
    \item $\Omega_{\rm Na\,I}$ also increases toward low redshift and broadly
    tracks $dN/dX$. The ratio $\Omega_{\rm Na\,I}/(dN/dX)$ shows no significant evolution, indicating little change in the mean measured \nai\ column density per absorber.
    
    \item Preliminary host-galaxy associations and quasar reddening measurements suggest that \nai\ absorbers trace dusty, dense neutral gas in and around galaxies. Dedicated galaxy surveys around \nai-bearing quasar sightlines, together with associated-ion measurements and simulations, will test this picture.
    
\end{itemize}

\section*{Data Availability}
The parent quasar catalog and spectra are publicly available as part of DESI DR1 at \url{data.desi.lbl.gov/doc/releases/dr1/}. The \nai~absorber catalog used in this work will be publicly available at \url{https://github.com/abhi0395/desi-dr1-nai}. The data used to reproduce the figures and statistical measurements presented in this work will be archived at \url{https://doi.org/10.5281/zenodo.22857887}. The quasar continuum-fitting code, \texttt{nmfqsofit}, is publicly available at \url{https://github.com/abhi0395/nmfqsofit}. The automated absorber-search code
used to construct the catalog is publicly available through the \texttt{qsoabsfind} package at \url{https://github.com/abhi0395/qsoabsfind}.

\section*{Acknowledgements}

We thank the referee for constructive comments that helped improve the paper. We also thank Vikram Khaire, Raghunathan Srianand, and Paryag Sharma for helpful comments and suggestions on the manuscript. All the computations were performed at the DOE's high-performance computing facility, NERSC at Berkeley Lab. This research used data obtained with the Dark Energy Spectroscopic Instrument (DESI). DESI construction and operations is managed by the Lawrence Berkeley National Laboratory. This material is based upon work supported by the U.S. Department of Energy, Office of Science, Office of High-Energy Physics, under Contract No. DE–AC02–05CH11231, and by the National Energy Research Scientific Computing Center, a DOE Office of Science User Facility under the same contract. Additional support for DESI was provided by the U.S. National Science Foundation (NSF), Division of Astronomical Sciences under Contract No. AST-0950945 to the NSF’s National Optical-Infrared Astronomy Research Laboratory; the Science and Technology Facilities Council of the United Kingdom; the Gordon and Betty Moore Foundation; the Heising-Simons Foundation; the French Alternative Energies and Atomic Energy Commission (CEA); the National Council of Humanities, Science and Technology of Mexico (CONAHCYT); the Ministry of Science and Innovation of Spain (MICINN), and by the DESI Member Institutions: \url{www.desi.lbl.gov/collaborating-institutions}. The DESI collaboration is honored to be permitted to conduct scientific research on I’oligam Du’ag (Kitt Peak), a mountain with particular significance to the Tohono O’odham Nation. Any opinions, findings, and conclusions or recommendations expressed in this material are those of the author(s) and do not necessarily reflect the views of the U.S. National Science Foundation, the U.S. Department of Energy, or any of the listed funding agencies.

\section*{Disclosure of Generative AI Usage}

We used ChatGPT-5.6-Sol to assist with grammar and language editing of the manuscript. It was also used in a limited capacity to improve code documentation, clean up existing scripts, and optimize some computational steps. All such code changes were reviewed and validated by the lead author. The scientific data, methods, interpretation of the results, and conclusions presented in this work are entirely those of the authors.

\software{\texttt{Matplotlib}
  \citep{hunter07a}, \texttt{NumPy} \citep{harris20a}, \texttt{SciPy} \citep{virtanen20a}, \texttt{Astropy} \citep{astropy2013,astropy2018,astropy2022}, 
  \texttt{qsoabsfind} \citep{anand2021, qsoabsfind2025}, 
  \texttt{nmfqsofit} \citep{anand2021},
  \texttt{numba} \citep{numba2015}}

\appendix

\section{Completeness and Purity Analysis}
\label{appendix:selection_function}

\begin{figure*}
    \centering
    \includegraphics[width=0.46\textwidth]{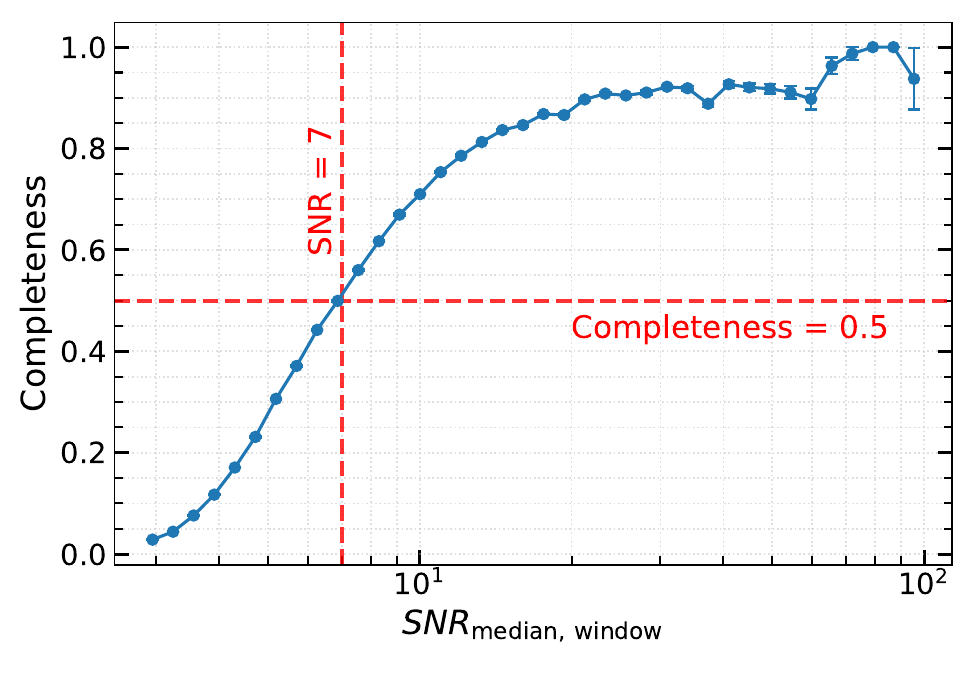}
    \hfill
    \includegraphics[width=0.46\textwidth]{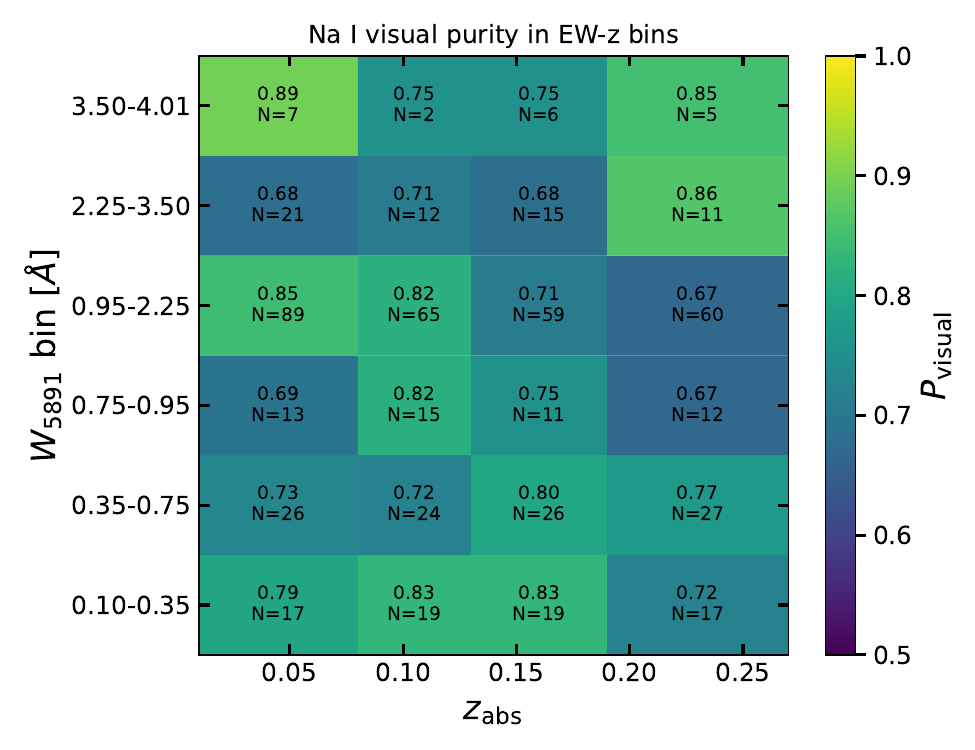}
    \caption{Completeness and purity diagnostics for our \nai\ absorber catalog. \textbf{Left:} Injection-recovery completeness as a function of the median per-pixel S/N in the \nai\ search window. The vertical dashed line marks the adopted threshold, ${\rm S/N}_{\rm median,window}=7$, and the horizontal dashed line marks 50\% completeness. \textbf{Right:} Visual purity in bins of absorber redshift and rest-frame equivalent width. Each cell shows the mean visual-purity score, $P_{\rm visual}$, and the number of visually inspected candidates in that bin. The inspected candidates were selected through stratified sampling in $(W_{5891}, z_{\rm abs})$, with visual purity estimated separately in each 2D bin. These binned purity values are used in the statistical weights together with the injection-recovery completeness correction.}
    \label{fig:selection_function_diagnostics}
\end{figure*}

Figure~\ref{fig:selection_function_diagnostics} shows two diagnostics used
to define and correct the statistical sample. The left panel shows the
injection-recovery completeness as a function of the median per-pixel S/N
in the \nai\ search window, averaged over all injected absorbers across the
simulated equivalent-width range. The completeness rises rapidly with
spectral quality, reaching $\simeq50\%$ near
${\rm S/N}_{\rm median,window}\simeq7$. We therefore adopt
${\rm S/N}_{\rm median,window}\geq7$ for the final parent QSO sample used
to search for \nai\ absorbers. The small number of objects at very high S/N does not significantly affect this trend: fewer than $0.5\%$ of the parent QSOs have
${\rm S/N}_{\rm median,window}>30$. Nearly all of these are very bright
quasars, typically with $r<17.5$.

The right panel shows the visual-purity estimate in bins of absorber
redshift and rest-frame equivalent width, $W_{\rm 5891}$. The binned values $P_{\rm visual}(W_{5891},z_{\rm abs})$ are assigned to absorbers in the corresponding bins when computing the purity- and completeness-corrected statistical measurements.

\section{Median Composite Spectra in the Absorber Frame}
\label{appendix:residual_stacks}

As an additional check on the detected \nai\ systems, we construct
absorber-frame composites of the continuum-normalized spectra in
bins of \nai\ equivalent width. For each absorber, the spectrum is
shifted to the absorber rest frame,
$\lambda_{\rm abs}=\lambda_{\rm obs}/(1+z_{\rm abs})$, and interpolated onto a common wavelength grid with the same spacing as the DESI spectra
($\Delta\lambda=0.8\,\Ang$). The spectra are then combined using the median at each wavelength, which reduces the influence of outliers and unrelated absorption features. For the redshift range of our \nai\ sample, the corresponding \caii\ H\&K transitions remain within the DESI wavelength coverage for essentially all absorbers.

\begin{figure*}
    \centering
    \includegraphics[width=0.46\textwidth]{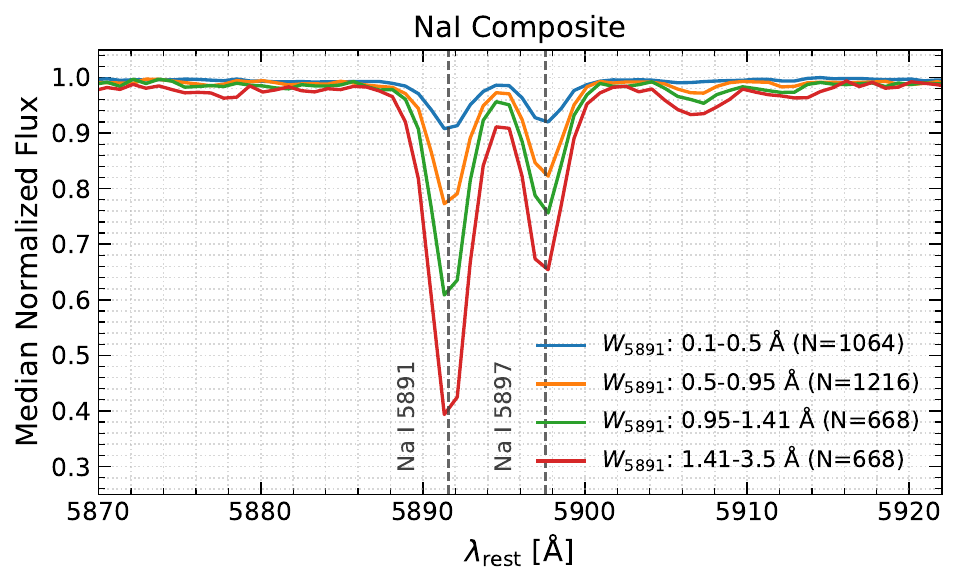}
    \hfill
    \includegraphics[width=0.46\textwidth]{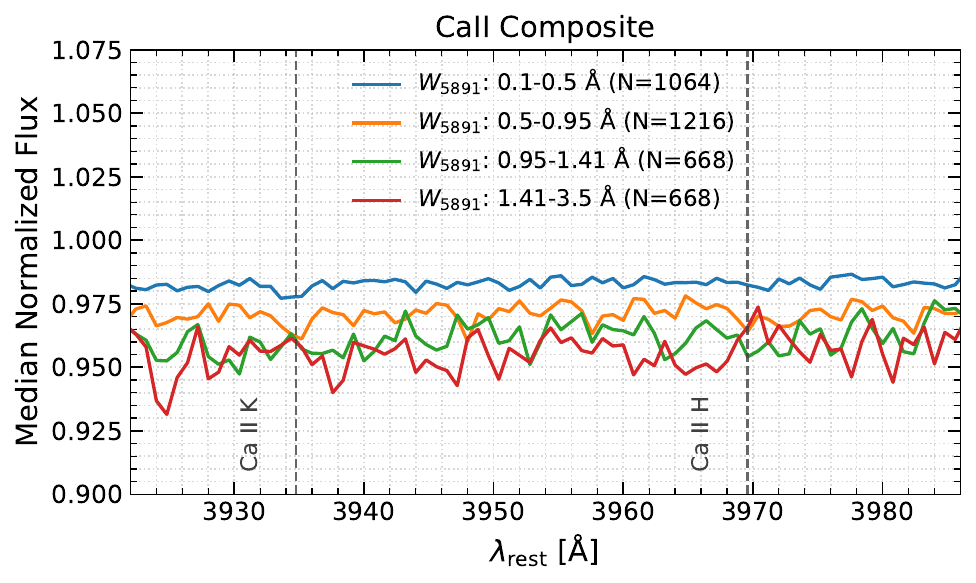}
    \caption{Absorber-frame median stacks of the continuum-normalized spectrum spectra in bins of $W_{5891}$. \textbf{Left:} Stacks around the \nai\ D doublet. Vertical dashed lines mark the expected positions of the two transitions. The coherent absorption and its increasing strength with $W_{5891}$ provide an independent visual check of the catalog. \textbf{Right:} Stacks around the \caii\ H\&K transitions, where no clear median absorption is detected. The $20$ systems with $W_{5891}>3.5\,\Ang$ are not included in the stacks, since they lie outside the adopted fiducial equivalent-width range and are
    too few to define a statistically useful additional bin.}
    \label{fig:nai_caii_residual_stack}
\end{figure*}

Figure~\ref{fig:nai_caii_residual_stack} shows the resulting stacks around the \nai\ D doublet (left panel) and the \caii\ H\&K transitions (right panel). The \nai\ stacks show clear absorption at the doublet wavelengths in all equivalent-width bins. The absorption strength increases from the weakest to the strongest $W_{5891}$ bin, consistent with the catalog measurements. This provides an independent visual check that the detected systems align coherently in the absorber rest frame.

We measure the \nai\ equivalent widths directly from the median stacks. The integration windows are determined from double-Gaussian fits to the stacked profiles. Uncertainties are estimated by bootstrap resampling the absorbers and repeating the stacking and equivalent-width measurements 500 times. The resulting measurements are listed in Table~\ref{tab:stack_ews}.

\begin{table}
\centering
\caption{Equivalent-width measurements from the median absorber-frame composite spectra. The uncertainties on the \nai\ equivalent widths are estimated from 500 bootstrap realizations. For \caii, the last two columns give the $3\sigma$ equivalent-width sensitivities after local continuum renormalization.}
\label{tab:stack_ews}
\begin{tabular}{ccccc}
\hline
$W_{5891}$ bin
& $N_{\rm abs}$
& $W_{5891}^{\rm stack}$
& $W_{5897}^{\rm stack}$
& $3\sigma(W_{3934})$ \quad $3\sigma(W_{3969})$ \\
(\Ang) & & (\Ang) & (\Ang) & (\Ang) \\
\hline
0.10--0.50  & 1064 & $0.295\pm0.004$ & $0.234\pm0.003$ & $<0.011 \quad <0.011$ \\
0.50--0.95  & 1216 & $0.671\pm0.006$ & $0.485\pm0.007$ & $<0.018 \quad <0.023$ \\
0.95--1.41  & 668  & $1.110\pm0.013$ & $0.687\pm0.012$ & $<0.035 \quad <0.036$ \\
1.41--3.50  & 668  & $1.914\pm0.046$ & $1.039\pm0.043$ & $<0.039 \quad <0.040$ \\
0.1--3.50  & 3616  & $0.702\pm0.011$ & $0.471\pm0.006$ & $<0.01 \quad <0.01$ \\
0.1--4.0  & 3636  & $0.708\pm0.011$ & $0.473\pm0.006$ & $<0.01 \quad <0.01$ \\
\hline
\end{tabular}
\end{table}

The corresponding \caii\ stacks do not show a significant absorption feature
at the expected H\&K wavelengths in any of the four $W_{5891}$ bins. No clear
\caii\ signal is detected even when all \nai\ absorbers are combined into a
single stack. The local residual level around \caii\ shows broad offsets from unity,
particularly in the higher-$W_{5891}$ bins. Such offsets can arise even with
an unbiased continuum normalization, since unrelated absorption along the
individual sightlines is smeared in the \nai\ absorber rest frame. We
therefore renormalize each \caii\ stack using a local linear continuum while
masking the expected H\&K line positions. Equivalent widths are then measured
over fixed windows around the two transitions. Their uncertainties are
estimated using the same bootstrap procedure adopted for the \nai\
measurements. The resulting $3\sigma$ equivalent-width sensitivities are
listed in Table~\ref{tab:stack_ews}.

The weak \caii\ signal may reflect differences in ionization or depletion
between \nai- and \caii-bearing gas, although sensitivity and residual
continuum systematics may also contribute. This is broadly consistent with
measurements in the Galactic ISM. \citet{welty1996} find that, over the
\nai\ column-density range relevant to our absorbers,
$\mathcal{N}(\mathrm{Na\,I})/\mathcal{N}(\mathrm{Ca\,II})$ is typically of order
$\sim10-50$, with substantial scatter and still larger ratios for some of
the strongest \nai\ components. If a similar relation applies here, much
weaker \caii\ absorption than \nai\ would be expected.

\section{Robustness of the incidence evolution}
\label{appendix:robustness}

\begin{figure*}
    \includegraphics[width=0.475\linewidth]{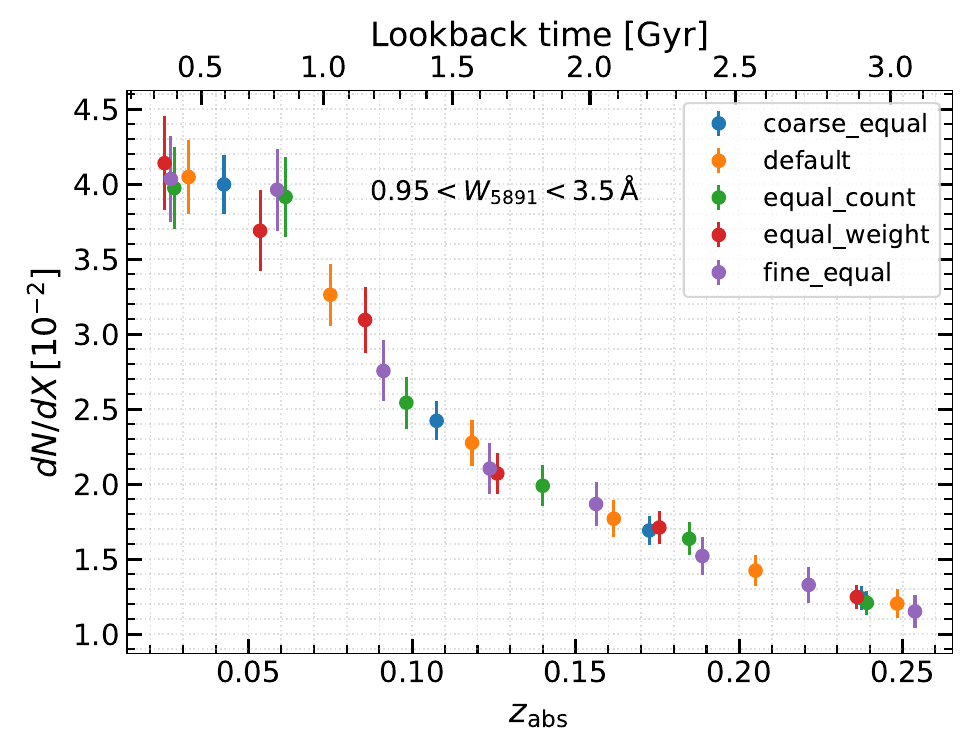}
    \includegraphics[width=0.475\linewidth]{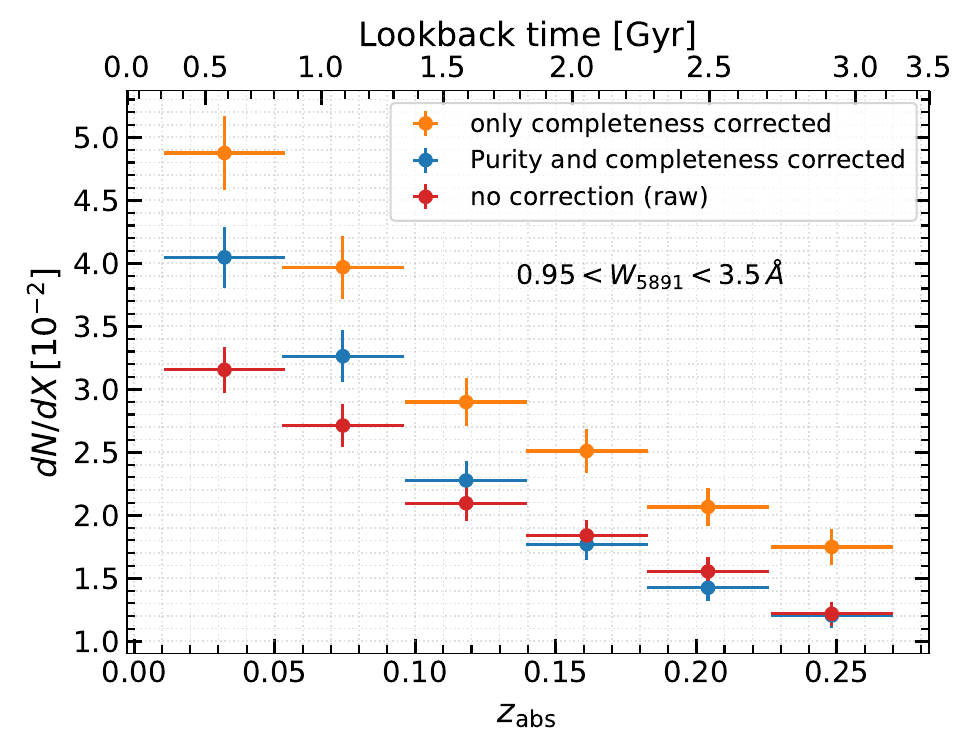}
    \caption{\textbf{Left:} Robustness of the \nai\ incidence to the adopted redshift binning. The $dN/dX$ measurements are repeated with several binning schemes using the same fiducial weights as in the main analysis. The increase toward low redshift is preserved in all cases. \textbf{Right:} Sensitivity of $dN/dX$ to the weighting scheme. The fiducial purity- and completeness-corrected measurements are compared with alternative weights. The normalization changes slightly, but the redshift trend remains unchanged.}
\label{fig:dndx_robustness}
\end{figure*}

Here, we test whether the observed evolution in the \nai\ incidence depends on the choices made in the fiducial analysis. The left panel of
Figure~\ref{fig:dndx_robustness} shows $dN/dX$ measured with alternative redshift binnings. Although the individual values vary slightly with the bin boundaries, the rise toward low redshift is preserved for the fiducial sample. This indicates that the inferred evolution is not sensitive to the particular binning used in the main analysis.

The right panel of Figure~\ref{fig:dndx_robustness} shows the sensitivity of $dN/dX$ to the statistical weights applied to individual absorbers. We compare the fiducial purity- and completeness-corrected measurements with results obtained using alternative weighting schemes. The normalization changes modestly, but the redshift trend and the rise toward the present epoch remain qualitatively unchanged. We therefore conclude that the main evolutionary result is robust to both the redshift binning and weighting scheme.

\bibliographystyle{aasjournal}
\bibliography{refs}

\end{document}

%% file: authors.tex

\author[0000-0003-2923-1585]{Abhijeet~Anand}
\affiliation{Inter-University Centre for Astronomy \& Astrophysics, Post Bag 04, Pune, India 411007}
\affiliation{Lawrence Berkeley National Laboratory, 1 Cyclotron Road, Berkeley, CA 94720, USA}
\email[show]{abhijeet.anand@iucaa.in}

\author[0000-0003-3938-8762]{Sowgat~Muzahid}
\affiliation{Inter-University Centre for Astronomy \& Astrophysics, Post Bag 04, Pune, India 411007}
\email[show]{sowgat@iucaa.in}